\documentclass[11pt]{article}
\usepackage{amssymb}
\usepackage{graphicx}
\usepackage{theorem}
[1999/03/29 PSNFSS v.7.2
Times font as default roman
: S Rahtz]

\newcommand \be  {\begin{equation}}
\newcommand \bea {\begin{eqnarray}}
\newcommand \ee  {\end{equation}}
\newcommand \eea {\end{eqnarray}}
\newcommand \E {{\rm E}}
\newcommand \cov{{\rm Cov}}
\newcommand \var{{\rm Var}}
\newcommand \A {{\cal A}}
\newcommand \B {{\cal B}}
\newcommand \inA { \in {\cal A}}
\newcommand \inB { \in {\cal B}}
\renewcommand \v { {\tilde v}}

\theoremheaderfont{\scshape}
\theoremstyle{break}
\theorembodyfont{\itshape}

\newtheorem{lemma}{Lemma}

\begin{document}

\title{Investigating Extreme Dependences: Concepts and Tools
\footnote{We acknowledge helpful discussions and
exchanges with J.P. Laurent, F. Lindskog and V. Pisarenko. This work 
was partially
supported by
the James S. Mc Donnell Foundation 21st century scientist award/studying
complex system.}}
\author{Y. Malevergne$^{1,2}$ and D. Sornette$^{1,3}$ \\
\\
$^1$ Laboratoire de Physique de la Mati\`ere Condens\'ee CNRS UMR 6622\\
Universit\'e de Nice-Sophia Antipolis, 06108 Nice Cedex 2, France\\
$^2$ Institut de Science Financi\`ere et d'Assurances - Universit\'e Lyon I\\
43, Bd du 11 Novembre 1918, 69622 Villeurbanne Cedex\\
$^3$ Institute of Geophysics and Planetary Physics
and Department of Earth and Space Science\\
University of California, Los Angeles, California 90095, USA\\
\\
email: Yannick.Malevergne@unice.fr and sornette@unice.fr\\
fax: (33) 4 92 07 67 54\\
}

\maketitle

\begin{abstract}

We investigate the relative information content of
six measures of dependence between two random variables $X$ and $Y$
for large or extreme events for several models of interest for
financial time series.
The six measures of dependence are
respectively the linear correlation
$\rho^+_v$ and Spearman's rho $\rho_s(v)$ conditioned on signed
exceedance of one variable
above the threshold $v$,
or on both variables ($\rho_u$), the linear correlation $\rho^s_v$ conditioned
on absolute value exceedance (or large volatility) of one variable,
the so-called asymptotic tail-dependence $\lambda$ and a probability-weighted
tail dependence coefficient ${\bar \lambda}$. The models are
the bivariate Gaussian distribution, the bivariate Student's distribution,
and the factor model for various distributions of the factor.
We offer explicit analytical formulas as well as numerical
estimations for these six measures of dependence
in the limit where $v$ and $u$ go to infinity.
This provides a quantitative proof that conditioning on exceedance leads to
conditional correlation coefficients that may be very different from
the unconditional
correlation and gives a straightforward mechanism for fluctuations or
changes of
correlations, based on fluctuations of volatility or changes of trends.
Moreover, these various measures of dependence exhibit
different and sometimes opposite behaviors,
suggesting  that, somewhat similarly to risks whose adequate
characterization requires an extension
beyond the restricted one-dimensional
measure in terms of the variance (volatility)
to include all higher order cumulants or more generally the knowledge
of the full
distribution, tail-dependence has also a multidimensional character.

\end{abstract}

%*******************************
%*         Introduction        *
%*******************************

\section*{Introduction}

The Oct. 19, 1987, stock-market crash stunned Wall Street professionals, hacked
about \$1 trillion off the value of all U.S. stocks, and elicited
predictions of
another Great Depression. On ``Black Monday,'' the Dow Jones industrial average
plummeted $508$ points, or $22.6$ percent, to $1,738.74$. It was both
the largest
one-day point and percentage loss ever for the blue-chip index until the
large loss of 684.81 points due to the events on September 11, 2001. The
broader markets followed the Dow downward. The S\&P 500 index lost more than
$20$ percent, falling $57.86$ to $224.84$. The Nasdaq Composite index dived
$46.12$ to $360.21$. No Dow
components emerged unscathed from Black Monday. Even market stalwarts suffered
massive share losses. IBM shed $31-3/4$ to close at $103-1/3$, while
USX lost $12-1/2$
to $21-1/2$ and Eastman Kodak fell $27-1/4$ to $62-7/8$. The crash splattered
technology stocks as well. On the Nasdaq, Apple Computer lost
$11-3/4$ to close at
$36-1/2$, while Intel dropped $10$ to $42$. Stocks descended quickly
on Black Monday,
with the Dow falling $200$ points soon after the opening bell to
trade at around
$2,046$. Yet by 10 a.m., the index had crept back up above $2,100$, beginning a
pattern of rebound and retreat that would continue for most of the day. Later,
with 75 minutes left in the trading day, it looked like the Dow would
escape with
a loss of ``only'' about $200$ points. But the worst was yet to come.
Starting at
about 2:45 p.m., a massive sell-off began, eventually ripping $300$
more points off
of the Dow. At the closing bell, the Dow appeared to have suffered an amazing
loss of about $400$ points. However, heavy volume kept the New York Stock
Exchange's computers running hours behind trading. Only about two hours later
would investors realize that the day's total loss exceeded $500$ points.

This event epitomizes the observation often reported by
market professionals that, ``during major
market events, correlations change dramatically'' \cite{Bookstaber}.
The possible existence of changes of correlation, or more precisely of
changes of dependence,
between assets in different market phases has obvious implications
in risk assessment, portfolio management and in the way policy and
regulation should be
performed.

There are two distinct classes of mechanisms for understanding
``changes of correlations'', not necessarily mutually exclusive.
\begin{itemize}
\item  It is possible that there are genuine changes with time of the
unconditional correlations and thus of the underlying
structure of the dynamical processes, as observed by identifying
shifts in ARMA-ARCH/GARCH processes
\cite{Granger}, in regime-switching models \cite{AB00,AC01}
or in contagion models \cite{Quintos1,Quintos2}.
\cite{LS95,TY99} and many others have shown that the hypothesis of a
constant conditional
correlation for stock returns or international equity returns must
be rejected. In fact, there is strong evidence that the correlations are not
only time dependent but also state dependent. Indeed, as shown by \cite{KW90,
RS98}, the correlations increase in periods of large volatility. Moreover,
\cite{LS01} have proved that the correlations across international equity
markets are trend dependent.

\item In contrast, a second class of explanation is that correlations
between two variables
conditioned on signed exceedance (one-sided) or on absolute value (volatility)
exceedance of one or both variables may
deviate significantly from the unconditional correlation
\cite{Boyer_etal,Loretan,Loretanenglish}. In other words,
with a fixed unconditional correlation $\rho$, the measured
correlation conditioned
of a given bullish trend, bearish trend, high or low market
volatility, may in general
differ from $\rho$ and be a function of the specific market phase. According
to this explanation, changes of correlation may be only a fallacious
appearance that stems from
a change of volatility or a change of trend of the market and not from a real
change of unconditional correlation. 
\end{itemize}

The existence of the second class of explanation
is appealing by its parsimony, as it posits that observed ``changes
of correlation''
may simply result from the way the measure of dependence is performed.
Therefore, before invoking additional mechanisms involving genuine changes of
unconditional dependence, it is highly desirable to characterize the
different possible ways with which higher or lower conditional
dependence can occur in
models with constant unconditional dependence. In order to make progress,
it is necessary to first distinguish between the different measures of
dependence between two variables for large or extreme events
that have been introduced in the literature, because the conclusions that one
can draw about the variability of dependence are sensitive to the choice
of its measure. These measures include
\begin{enumerate}

\item  the correlation conditioned on signed exceedance of one
or both
variables \cite{Boyer_etal,Loretan,Loretanenglish,Cizeau_etal},
that we call respectively $\rho^+_v$ and $\rho_u$, where $u$ and $v$
denote the thresholds above which the exceedances are calculated,

\item the correlation conditioned
on absolute value exceedance (or large volatility), above the
threshold $v$, of one or both variables
\cite{Boyer_etal,Loretan,Loretanenglish,Cizeau_etal}, that we call
$\rho^s_v$ (for a condition of exceedance on one variable),

\item the tail-dependence parameter $\lambda$, which has
a simple analytical expression when using copulas \\ \cite{EMS01,L99} such as
the Gumbel copula \cite{LS01},

\item the spectral measure associated with the tail index (assumed
to be the same of all assets) of extreme value multivariate
distributions \cite{Davis_etal,Star99,Muller},

\item tail indices of extremal correlations defined as
the upper or lower correlation of exceedances of ordered log-values
\cite{Quintos1},

\item confidence weighted forecast correlations \cite{Bhansali} or algorithmic
complexity measures \cite{Mansilla}.

\end{enumerate}

Our contribution to the literature is to provide explicit analytical
expressions
for several measures (the conditional correlation coefficients $\rho^+_v$,
$\rho^s_v$,  $\rho_u$, the conditional Spearman's rho $\rho_s(v)$ and the tail
dependence coefficients $\lambda$ and $\bar \lambda$) of dependence between two
variables for large or extreme events for several models of interest for
financial time series. These models are the bivariate Gaussian distribution,
the bivariate Student's distribution, and the one factor model for various
distributions of the factor. Initially, we hoped to show the existence of
logical links between some of these measures, such as a vanishing
tail-dependence parameter $\lambda$ implies vanishing asymptotic $\rho^+_{v \to
+\infty}$ and $\rho_{u \to +\infty}$ or $\rho_s(v \to +\infty)$. We show that
this turns out to be wrong and one can construct simple examples for which all
possible combinations such as for instance ($\lambda=0$ and $\rho^+_{v \to
+\infty}=0$), ($\lambda=0$ and $\rho^+_{v \to +\infty} \neq 0$),
($\lambda \neq 0$ and $\rho^+_{v \to +\infty}=0$) and
($\lambda \neq 0$ and $\rho^+_{v \to +\infty} \neq 0$) occur.
Therefore, each of
these measures probe
a different quality of the dependence between two variables for large
or extreme events.
In addition, even if $\rho^+_{v \to +\infty}=0$ and $\rho_{u \to +\infty}$ are
zero, they decay in general extremely slowly as inverse powers of $v$
and $u$ and may thus
remain significant for most pratical applications.
Somewhat similarly to risks whose adequate characterization requires
an extension
beyond the restricted one-dimensional
measure in terms of the variance (volatility)
to include all higher order cumulants or more generally the knowledge
of the full
distribution \cite{Soretal1,Soretal2,Soretal3}, our results
suggest that tail-dependence has also a multidimensional
character.

One of our goal is also to provide a link
between two possible descriptions of financial data. The first one, based
on a purely statistical approach, is the multivariate or copula analysis,
while the second one, which relies on a principal components
analysis, is built on the
factor models. As we shall see, the dependence between two variables
for large or extreme events
may be quite different in these different models with the same
margins.

Section 1 describes three conditional correlation coefficients, namely
the correlation $\rho^+_v$ conditioned on signed exceedance of one variable,
or on both variables ($\rho_u$) and the correlation $\rho^s_v$ conditioned
on absolute value exceedance (or large volatility) of one variable.
\cite{Boyer_etal} have already provided the general expression of $\rho^+_v$
and $\rho^s_v$ for the Gaussian bivariate model. We use this result
to give their
$v$ dependence for large $v$: $\rho^+_v \propto 1/v$ and $1-\rho^s_v
\propto 1/v^2$.
We also give an intuitive explanation of these results. We then
provide the general expression of $\rho^+_v$
and $\rho^s_v$ for the Student's bivariate model with $\nu$ degrees
of freedom. We show that both $\rho^+_v$ and $\rho^s_v$ go to a
constant for large $v$.
For the factor model $X=\alpha Y + \epsilon$, we provide a
general expression of the conditional correlation coefficient whatever
the distributions of $Y$ and $\epsilon$ may be. In the special case when
$Y$ has a Student's distribution, we find
$1-\rho^+_v \propto 1-\rho^s_v \propto 1/v^2$.
Conditioning on both variables, we are able to provide the asymptotic
dependence of
$\rho_u$ only for the bivariate Gaussian model: $\rho_u \propto
1/u^2$  for $u \to +\infty$.

In section 2, to account for the deficiencies of the correlation
coefficient, we propose an alternative measure of dependence, the
conditional Spearman's rho, which is related to the probability of
concordance and discordance of several events drawn from the same
probability distribution. This measure provides an important
improvement with respect to the correlation coefficient since it only
takes into account the dependence structure of the variable and is not
sensitive to the marginal behavior of each variable. We perform numerical
computations to derive the behavior of the conditional Spearman's rho,
denoted by $\rho_s(v)$. This allows us to prove that there is no direct
relation between the Spearman's rho conditioned on large values
and the correlation coefficient conditioned on the same values. Therefore,
each of these coefficients quantifies a different kind of extreme
dependence.

Section 3 discusses the tail-dependence parameters $\lambda$ and $\bar
\lambda$. We first
recall their
definitions and values for Gaussian and Student's bivariate
distributions of $X$ and $Y$, already
known in the literature. For the Gaussian factor model, it is trivial to show
that $\lambda = 0$. A non-trivial result is obtained for the
Student's factor
model $X=\alpha Y + \epsilon$ where $Y$ and $\epsilon$ mutually independent:
$\lambda$ is found non-zero and a function only of $\alpha$ and of
the the scale factor of $\epsilon$. More generally, a theorem
established in \cite{MS02} allows one to calculate the coefficient of
tail dependence for any distribution of the factor and shows that
$\lambda$ vanishes for any rapidly varying factor.

Section 4 provides a synthesis and comparison between these different results.
A first important message is that there is no unique measure of extreme
dependence. Each of the coefficients of extreme dependence that we have studied
provides a specific quantification that is sensitive to a certain 
combination of
the marginals and of the copula of the two random variables.
Similarly to risks whose
adequate  characterization requires an extension beyond the restricted
one-dimensional measure in terms of the variance (volatility)
to include  the knowledge of the full distribution, tail-dependence
has also a multidimensional character. A second important message
is that the increase of some of the coefficients of tail dependence as
one goes more in the tails does not necessarily signals a genuine increase
of the unconditional correlation or dependence between the two variables. Our
calculations firmly confirm that this increase is a general and unvoidable
result of the statistical properties of many multivariate models of dependence.

%*******************************
%                   Section 1              *
%*******************************

\section{Conditional correlation coefficient}

In this section we discuss the properties of the correlation coefficient
conditioned on one variable. We study the difference when conditioning
on the signed values or on absolute values\footnote{Conditioning on absolute
value of the variable of interest is only meaningful when its distribution is
symetric.} of the variable. This allows us to conclude that conditioning on
signed values is generally more efficient than conditioning on absolute values,
and that, as already underlined by \cite[for instance]{Boyer_etal}, the
conditional correlation coefficient suffers from a bias which forbides its use
as a tool to measure a change in the correlation between two assets when the
volatility increases, as seen in many papers about contagion.

We then present some empirical illustrations of the evolution of the
correlation between an asset and the market portfolio. We conclude this section
with some considerations about the correlation coefficient conditioned on both
variables which allows us to assert that this kind of conditioning does not
provide a real improvement with respect to the correlation coefficient
conditioned on one signed variable.

\subsection{Definition}
We study the
correlation coefficient $\rho_\A$ of two real random variables $X$ and $Y$
conditioned on $Y \inA$, where $\A$ is a
subset of $\mathbb R$ such that $\Pr \{Y \inA \} >0$.

By definition, the conditional correlation coefficient $\rho_\A$ is given by
\be
\label{eq:rho_a}
\rho_\A=\frac{\cov(X, Y ~|~ Y \inA)
}{\sqrt{\var(X ~|~ Y\inA) \cdot  \var(Y ~|~ Y \inA)}}~.
\ee

Applying this general expression of the conditional correlation coefficient, we
will give closed formulae for several standard distributions and models.

\subsection{$X$ and $Y$ have a Gaussian distribution}
\label{sec:gauss}

Let the variables $X$ and $Y$ have a multivariate Gaussian
distribution with (unconditional) correlation coefficient $\rho$. The following
result have been proved by \cite{Boyer_etal}~:
\be
\rho_\A = \frac{\rho}{\sqrt{\rho^2 + (1-\rho^2)
\frac{\var(Y)}{\var(Y~|~Y \inA)}}}~.
\label{jgbnkaq}
\ee
We can note that $\rho$ and $\rho_\A$ have the same sign, that $\rho_\A=0$ if
and only if $\rho=0$ and that $\rho_\A$ does not depend directly on $\var (X)$.
Note also that $\rho_\A$ can be either greater or smaller than $\rho$
since $\var(Y~|~Y\inA)$ can be either greater or smaller than
$\var(Y)$. We will
illustrate this property in the two following examples.

\subsubsection{Conditioning on $Y>v$}

Let the conditioning set be $\A = [v, +\infty)$, with $v \in {\mathbb
R_+}$. Thus
$\rho_\A$ is the correlation coefficient conditioned on $Y$ larger than $v$. It
will be denoted by $\rho_v^+$ in the sequel.
Assuming for simplicity, but without loss of generality, that
$\var(Y)=1$, we have
\be
\rho_v^+=
\frac{\rho}{\sqrt{\rho^2+\frac{1-\rho^2}{\var(Y~|~Y > v)}}}~.
\label{mgjrlkr}
\ee

Appendix \ref{app:os_ccc_gv} presents an exact calculation of
$\var(Y~|~ Y \inA)$, which leads, for large $v$, to
\be
\label{eq:os_ccc_gv}
\rho_v^+ \sim_{v \rightarrow \infty} \frac{\rho}{\sqrt{1-\rho^2}}\cdot
\frac{1}{v}~.
\ee
Thus, $\rho_v^+$ goes slowly to zero as $v$ goes to infinity.
Obviously, by symmetry, the conditional correlation coefficient
$\rho_v^-$, conditioned on $Y$ smaller than $v$, obeys the same formula.

\subsubsection{Conditioning on $|Y|>v$}
Let now the conditioning set be $\A =(- \infty, -v] \cup [v,
+\infty)$, with $v \in
{\mathbb R_+}$. Thus $\rho_\A$ is the correlation coefficient conditioned on
$|Y|$ larger than $v$, i.e., it is conditioned on large volatility of $Y$.
Still assuming $\var(Y)=1$, we denote it by $\rho_v^s$:
\be
\rho_v^s=
\frac{\rho}{\sqrt{\rho^2+\frac{1-\rho^2}{\var(Y~|~|Y| > v)}}}~.
\label{jgnnhklf}
\ee

Appendix \ref{app:os_ccc_gv2} presents the
calculation of $\var(Y~|~|Y|>V)$, which allows us to conclude that,
for large $v$,
\be
\label{eq:os_ccc_gv2}
\rho_v^s \sim_{v \rightarrow \infty}
\frac{\rho}{\sqrt{\rho^2+\frac{1-\rho^2}{2+v^2}}}~,
\ee
showing that $\rho_v^s$ goes to one as $v$ goes to infinity as
$1-\rho_v^s \sim_{v \rightarrow \infty} {1-\rho^2 \over \rho^2}~v^{-2}$.

\subsubsection{Conditioning on $Y>v$ versus $|Y|>v$}
In the case of two Gaussian random variables, the two conditional
correlation coefficients $\rho_v^+$ and $\rho_v^s$ thus exhibit
opposite behavior
since the conditional correlation coefficient $\rho_v^+$ is a decreasing
function of $v$ which goes to zero as $v \to +\infty$ while the
conditional correlation coefficient $\rho_v^s$ is an increasing function of
$v$ and goes to one as $v \to \infty$.

Let us provide an intuitive explanation (see also \cite{LS95}). As
seen from (\ref{mgjrlkr}),
$\rho_v^+$ is controlled by the dependence $\var(Y~|~Y > v) \propto 1/v^2$
derived in Appendix \ref{app:os_ccc_gv}. In contrast, as seen from
(\ref{jgnnhklf}),
$\rho_v^s$ is controlled by $\var(Y~|~|Y| > v) \propto v^2$
given in Appendix \ref{app:os_ccc_gv2}. The difference between $\rho_v^+$
and $\rho_v^s$ can thus be traced back to that between
$\var(Y~|~Y > v) \propto 1/v^2$ and $\var(Y~|~|Y| > v) \propto v^2$
for large $v$.

This results from the following effect.
For $Y>v$, one can
picture the possible realizations of $Y$ as those of a random
particle on the line,
which is strongly attracted to the origin by a spring (the Gaussian
distribution
that prevents $Y$ from
performing significant fluctuations beyond a few standard deviations)
while being forced to be on the right to a wall at $Y=v$. It is clear that the
fluctuations of the position of this particle are very small as it is strongly
glued to the unpenetrable wall by the restoring spring, hence the result
$\var(Y~|~Y > v) \propto 1/v^2$. In constrast, for the condition
$|Y|>v$, by the same argument, the fluctuations of the particle are hindered
to be very close to $|Y|=v$, i.e., very close to $Y=+v$ or $Y=-v$. Thus, the
fluctuations of $Y$ typically flip from $-v$ to $+v$ and vice-versa. It is thus
not surprising to find $\var(Y~|~|Y| > v) \propto v^2$.

This argument makes intuitive the results
$\var(Y~|~Y > v) \propto 1/v^2$ and $\var(Y~|~|Y| > v) \propto v^2$
for large $v$
and thus the results for $\rho_v^+$ and for $\rho_v^s$ if we use
(\ref{mgjrlkr})
and (\ref{jgnnhklf}). We now attempt to justify
$\rho_v^+ \sim_{v \rightarrow \infty} \frac{1}{v}$ and
$1-\rho_v^s \sim_{v \rightarrow \infty} 1/v^{2}$ directly by the following
intuitive argument. Using the picture of particles, $X$ and $Y$ can
be visualized
as the positions of two particles which fluctuate randomly. Their joint
bivariate Gaussian distribution with non-zero unconditional correlation amounts
to the existence of a spring that ties them together. Their Gaussian marginals
also exert a spring-like force attaching them to the origin. When $Y>v$, the
$X$-particle is teared off between two extremes, between $0$ and $v$. When
the unconditional correlation $\rho$ is less than $1$, the spring attracting to
the origin is stronger than the spring attracting to the wall at $v$.
The particle
$X$ thus undergoes tiny fluctuations around the origin that are
relatively less and less attracted by the $Y$-particle, hence the result
$\rho_v^+ \sim_{v \rightarrow \infty} \frac{1}{v} \to 0$.
In constrast, for $|Y|> v$, notwithstanding the still strong attraction of the
$X$-particle to the origin, it can follow the sign of the
$Y$-particle without paying too much cost in matching its amplitude $|v|$.
Relatively tiny fluctuation of the $X$-particle but of the same sign
as $Y \approx \pm v$
will result in a strong $\rho_v^s$, thus justifying that $\rho_v^s \to 1$ for
$v \to +\infty$.

\subsection{$X$ and $Y$ have a Student's distribution}

Let the variables $X$ and $Y$ have a multivariate Student's distribution with
$\nu$ degrees of freedom and an (unconditional) correlation coefficient $\rho$.
According to the proposition stated in appendix \ref{app:rho_std}, we have
\be
\rho_\A  = \frac{\rho}{\sqrt{\rho^2 + \frac{\E[
\E(X^2~|~Y)- \rho^2 Y^2~|~ Y\inA]}{\var(Y~|~Y\inA)}}}~.
\label{gnnnv}
\ee
Appendix \ref{app:rho_std}
gives the explicit formulas of $\E[\E(X^2~|~Y)- \rho^2
Y^2~|~ Y\inA]$ and $\var(Y~|~Y\inA)$.

Expression (\ref{gnnnv}) is the analog for a Student bivariate distribution
to (\ref{jgbnkaq}) derived above for the
Gaussian bivariate distribution.
Again, $\rho$ and $\rho_\A$ share the following properties: they have
the same sign,
$\rho_\A$ equals zero if and only if $\rho$ equals zero and
$\rho_\A$ can be either greater or smaller than $\rho$.
We now apply this general formula (\ref{gnnnv}) to the calculus of $\rho_v^+$
and $\rho_v^s$.

\subsubsection{Conditioning on $Y>v$}

We consider the conditioning set $\A = [v, +\infty)$ with $v
\in {\mathbb R_+}$. The expressions of
$\E[\E(X^2~|~Y)- \rho^2 Y^2~|~ Y\inA]$ and $\var(Y~|~Y\inA)$, in such a case,
can be obtained in closed form and are given in appendix \ref{app:os_ccc_sv}.
This allow us to conclude that, for large $v$,
\be
\rho_v^+ \longrightarrow \frac{\rho}{\sqrt{  \rho^2 + (\nu-1)
\sqrt{\frac{\nu-2}{\nu}}~ (1-\rho^2)}}~,
\label{mhytl}
\ee
which is a non vanishing contant, excepted for $\rho=0$. Moreover, for $\nu$
larger than $\nu_c \simeq 2.839$, this constant is smaller than the
unconditional correlation coefficient $\rho$, for all value of $\rho$.

\subsubsection{Conditioning on $|Y|>v$}

The conditioning set is now $\A=(- \infty, -v] \cup [v, +\infty)$, with $v
\in {\mathbb R_+}$. Appendix \ref{app:os_ccc_sv2} gives the closed
expressions of
$\E[\E(X^2~|~Y)- \rho^2 Y^2~|~ Y\inA]$ and $\var(Y~|~Y\inA)$.
For large $v$, this leads to
\be
\rho_v^s \longrightarrow \frac{\rho}{\sqrt{  \rho^2 + \frac{1}{(\nu-1)}
\sqrt{\frac{\nu-2}{\nu}}~ (1-\rho^2)}}~,
\label{kuiigjj}
\ee
which is again a non vanishing contant. Contrarily to the previous case, this
constant is alway larger than $\rho$, whatever $\nu$ (larger than two)  may be.

\subsubsection{Conditioning on $Y>v$ versus on $|Y|>v$}

The results (\ref{mhytl}) and (\ref{kuiigjj}) are valid for $\nu> 2$,
as one can expect since the second moment has to exist for
the correlation coefficient to be defined. We remark that here,
contrarily to the Gaussian case, the conditioning set is not really
important. Indeed with both conditioning set, $\rho_v^+$ and
$\rho_v^s$ goes to a constant different from zero and one, when $v$
goes to infinity. This striking difference can be explained by the
large fluctuations allowed by the Student's distribution, and can be
related to the fact that the coefficient of tail dependence for this
distribution does not vanish even though the variables are
anti-correlated (see section \ref{sec:ts} below).

Contrarily to the Gaussian distribution which binds the fluctuations of
the variables near the origin, the Student's distribution allows for
'wild' fluctuations. These properties are thus responsible for the
result that, contrarily to the Gaussian case for which the conditional
correlation coefficient goes to zero when conditioned on large signed values
and goes to one when conditioned on large unsigned values, the conditional
correlation coefficient for Student's variables have a similar
behavior  in both cases. Intuitively, the large fluctuations of $X$
for large $v$ dominate and control the asymptotic dependence.

\subsection{$X$ and $Y$ are related by a linear equation (factor model)}

We now assume that $X$ and $Y$  are two random variables following the equation
\be
X= \alpha Y + \epsilon~,
\label{mgmmsaa}
\ee
where $\alpha$ is a non random real coefficient and $\epsilon$ an idiosyncratic
noise independent of $Y$, whose distribution admits a centered moment of
second order $\sigma_\epsilon ^2$. Let us also denote by $\sigma_y^2$ the
second centered moment of the variable $Y$.
This kind of relation between $X$ and $Y$ is the so-called {\it one factor
model} whose applications in finance can be traced back to \cite{R76}.
This one factor model with independence between $Y$ and $\epsilon$ is of course
naive for concrete applications, as it neglects the potential
influence of other
factors in the determination of $X$.
But for our purpose, it provides a simple illustrative model with
rich and somewhat
surprising results.

Appendix \ref{app:rho_fac} shows that the conditional correlation
coefficient of $X$ and $Y$ is
\be
\label{eq:rho_fac}
\rho_\A = \frac{\rho}{\sqrt{\rho^2 + (1-\rho^2)
\frac{\var(y)}{\var(y~|~y\inA)}}}~,
\label{mjhjjjs}
\ee
where
\be
\rho = \frac{\alpha \cdot \sigma_y}{\sqrt{\alpha^2 \cdot
\sigma_y^2+\sigma_\epsilon^2}}
\label{mjgjjnsa}
\ee
denotes the unconditional correlation
coefficient of $X$ and $Y$. Note that the term $\sigma_{\epsilon}^2$
in the expression
(\ref{mjgjjnsa}) of $\rho$ is the only place where the influence of
the idiosynchratic
noise is felt.

Expression (\ref{mjhjjjs}) is the same as (\ref{jgbnkaq}) for the
bivariate Gaussian
situation studied in \ref{sec:gauss}. This is not surprising since, in the case
where $Y$ and $\epsilon$ have univariate Gaussian distributions, the joint
distribution of $X$ and $Y$ is a bivariate Gaussian distribution. The
new fact is that
this expression
(\ref{eq:rho_fac}) remains true whatever the distribution of $Y$ and
$\epsilon$, provided that their second moments exist.

We now present the asymptotic expression of $\rho_\A$
for $Y$ with a Gaussian
or a Student's distribution. Note that
the expression of $\rho_\A$  is simple enough to allow for exact calculations
for a larger class of distributions.

\subsubsection{$Y$ has a Gaussian distribution}
Let us assume that $Y$ has a Gaussian distribution, while the
distribution of $\epsilon$ can be everything (provided that
$\E[\epsilon^2]<\infty$). The variance $\var(Y~|~ Y>v)$
behaves like $1/v^2$ for $v$ large enough according to (\ref{jghhdj})
while $\var(Y~|~|Y|>v) \sim v^2$ according to (\ref{nygwbbs}).
This obviously yields the same results as these given by equations
(\ref{eq:os_ccc_gv}) and (\ref{eq:os_ccc_gv2}).

\subsubsection{$Y$ has a Student's distribution}
Let us assume that $Y$ has a Student's distribution. Both
$\var(Y~|~ Y>v)$ and $\var(Y~|~ |Y|>v)$ are proportional to
$v^2$ for $v$ large enough, according to (\ref{ngvcvgz}) and (\ref{nvbgxts}).
This leads to
\be
\rho_v^{+,s} \sim  \frac{1}{\sqrt{1+\frac{K}{v^2}}}~,
\label{mghhek}
\ee
where $K$ is a positive constant.
$\rho_v^{+,s}$ thus goes to $1$ as $v$ goes to infinity with
$1-\rho_v^{+,s} \propto 1/v^2$.

\subsection{Empirical evidence}

\begin{figure}
\begin{center}
\includegraphics[width=15cm]{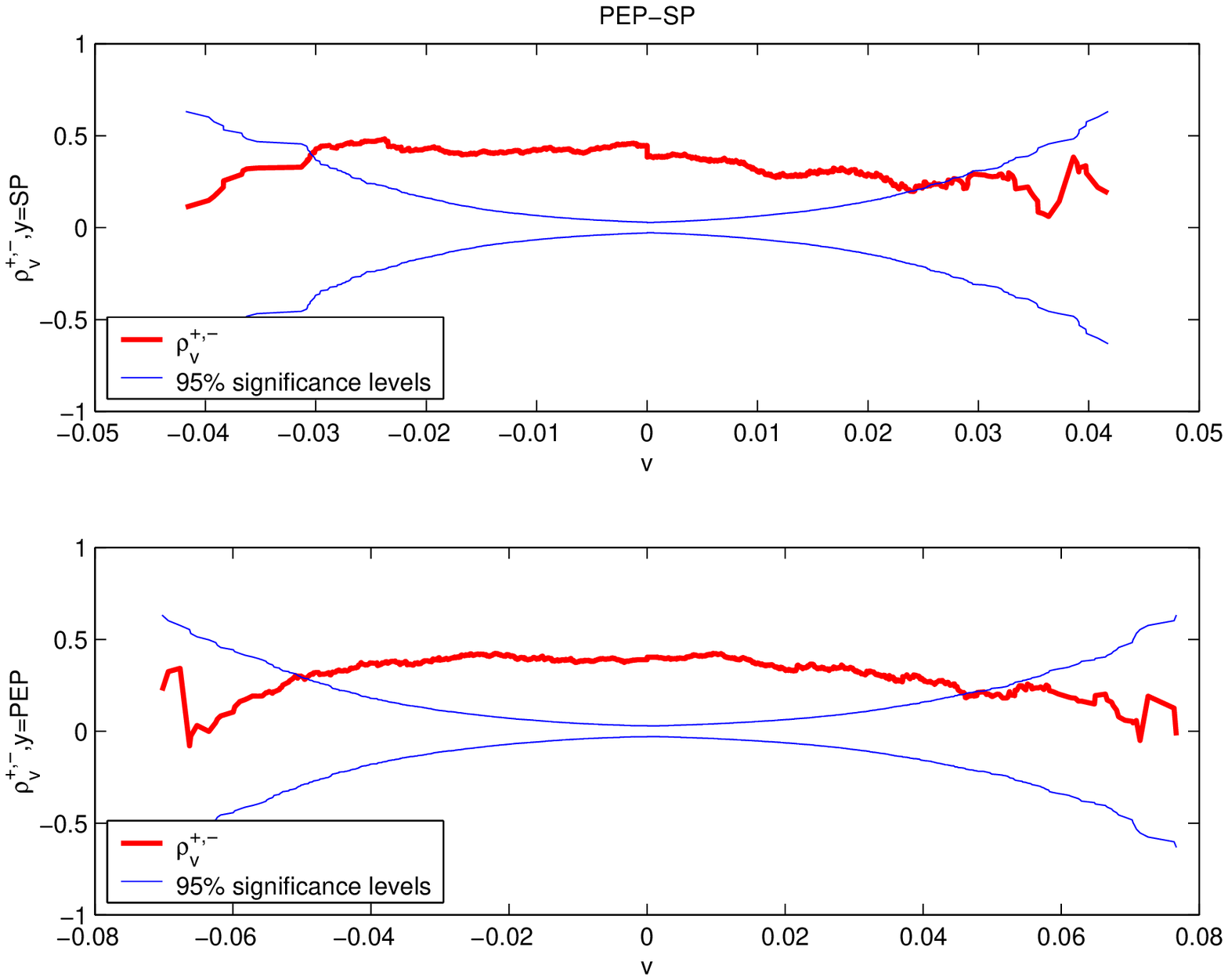}
\caption{\label{fig:corr_pep} In the upper panel, the thick curve depicts
the correlation coefficient between the Pepsico Incorporated daily returns and
the Standard \& Poor's 500 index daily returns conditional on the Standard \&
Poor's 500 index daily returns larger than (smaller than) a given positive
(negative) value $v$. The two thin curves represent the area within which we
cannot consider, at the 95\% confidence level, that the estimated correlation
coefficient is significantly different from zero. The lower panel gives the
same kind of information but for the correlation coefficient conditioned on the
Pepsico Incorporated daily returns larger than (smaller than) a given positive
(negative) value $v$.}
\end{center}
\end{figure}

Figures \ref{fig:corr_dis} and \ref{fig:corr_pep} show the
conditional correlation coefficient between the daily returns for the Disney
Company and the Standard \& Poor's 500 Index (fig \ref{fig:corr_dis}) and
between the daily returns of the Pepsico Incorporated and the Standard \&
Poor's 500 Index (fig \ref{fig:corr_pep}). For both figures, the upper panel
depicts the correlation coefficient conditional on the daily returns of the
Standard \& Poor's 500 Index larger than (resp. smaller than)  a given positive
(resp. negative) value $v$, while the lower panel gives the same
information when the conditioning is on the stock returns. The two thin
curves in each panel gives the values beyond which the estimated correlation
coefficient is significantly different from zero, at the 95\% confidence level.

We observe that
Pepsico and the Standard \& Poor's 500 Index shows no
significant correlation, neither in the upper nor in the lower tail,
whatever the conditioning variable may be. In contrast,
conditioned on large fluctuations of the Standard \& Poor's,
the correlation coefficient between Disney and the Standard \& Poor's
remains very significant. When conditioning on large negative
fluctuations of Disney, this correlation remains strong. However, for
large positive returns of Disney, the conditional correlation sinks
within the statistical uncertainty. This last feature is found more generally
for essentially all pairs of stocks and indices that we have
explored, for which the conditional correlation
is weak or non-existent for large positive returns. There is thus a
strong and significant asymmetry between the correlations conditioned 
on negative compared
to positive moves.

We stress that it would be incorrect to conclude that the correlation 
increases between
Disney and the Standard \& Poor's, based solely on the empirical 
determination of
the correlation coefficient conditioned on large returns of the 
Standard \& Poor's.
Indeed, we have shown by explicit analytical calculations that the
conditional correlations can increase when conditioning on larger fluctuations
without needing any variation of the unconditional correlation coefficient.

\begin{figure}
\begin{center}
\includegraphics[width=15cm]{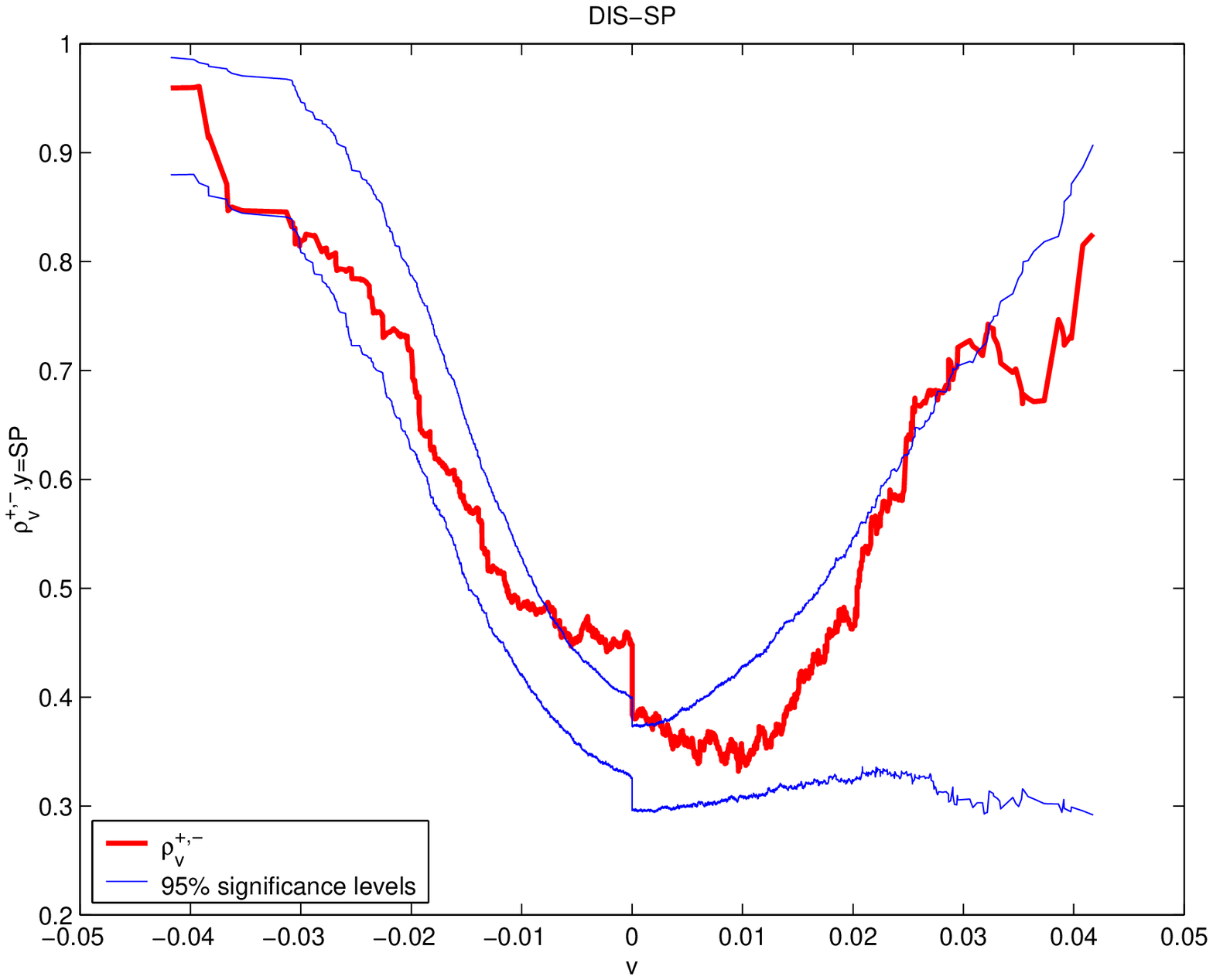}
\caption{\label{fig:boot_dis}  The thick curve depicts
the correlation coefficient between the Disney Company daily returns and the
Standard \& Poor's 500 index daily returns conditional on the Standard \&
Poor's 500 index daily returns larger than (smaller than) a given positive
(negative) value $v$. The two thin curves represent the borders within which
this conditional correlation coefficient must lay in order to comply, at the
95\% confidence level, with the assumption according to which
the Disney Company daily returns can be explained by a one factor model,
whose factor is given by the Standard \& Poor's 500 index daily returns.}
\end{center}
\end{figure}

\begin{figure}
\begin{center}
\includegraphics[width=15cm]{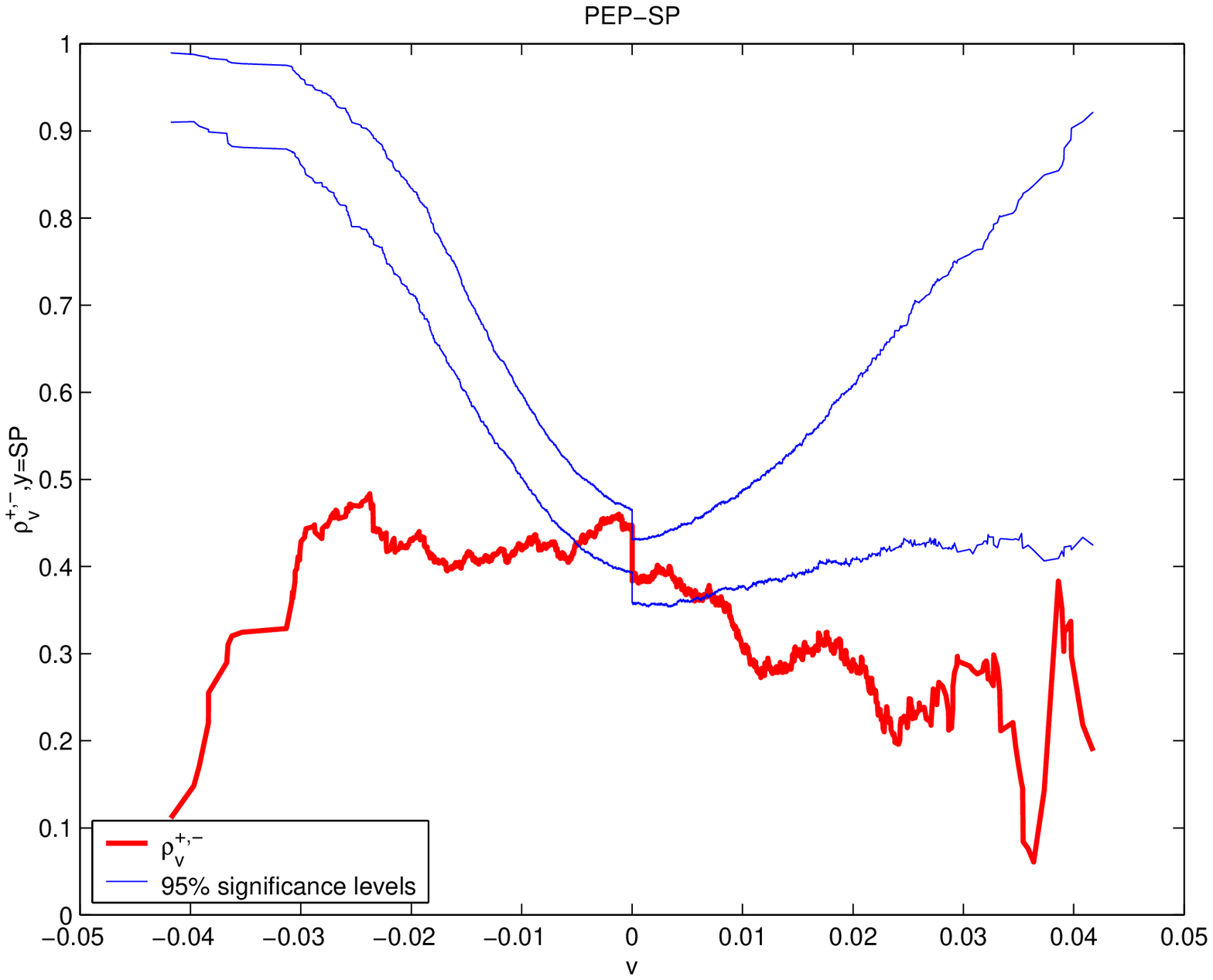}
\caption{\label{fig:boot_pep} the thick curve depicts
the correlation coefficient between the Pepsico Incorporated daily returns and
the Standard \& Poor's 500 index daily returns conditional on the Standard \&
Poor's 500 index daily returns larger than (smaller than) a given positive
(negative) value $v$. The two thin curves represent the borders within which
this conditional correlation coefficient must lay in order to comply, at the
95\% confidence level, with the assumption according to which
the Disney Company daily returns can be explained by a one factor model,
whose factor is given by the Standard \& Poor's 500 index daily returns.}
\end{center}
\end{figure}

Thus, the conditional correlations do not appear to be very useful 
tools for probing
the possible changes in the dependence structure between two assets. However,
coupled with the Capital Asset Pricing Model \cite{S64}, it provides 
interesting
insights. The CAPM predicts that the return of
any asset $X$ can be explained by the market return $Y$ according to the
relation
\be
\label{eq:fm}
X= \beta \cdot Y + \epsilon~,
\ee
where $\beta$ is a constant and $\epsilon$ is an idiosyncratic noise. As an
illustration, we will study the extreme dependence generated by this model and
compare it with the true observed dependence. Figure \ref{fig:boot_dis}
(respectively figure \ref{fig:boot_pep}) shows in thick line the correlation
coefficient between Disney (respectively Pepsico) and the Standard \& Poor's,
both conditioned on the Standard \& Poor's fluctuations. We have also performed
 the bootstrap estimation of  the 95\% confidence interval, represented
for the two thin lines in figures \ref{fig:boot_dis} and \ref{fig:boot_pep}, of
the conditional correlation coefficent predicted by the factor model
(\ref{eq:fm}).
The CAPM is found to provide a very good description
of the extreme dependence between Disney and the market, while it
completely fails to account for the lack of extreme dependence between Pepsico
and the market. The failure of the factor model to account for the conditional
correlation between Pepsico and the Standard \& Poor' s index may be due to
the impact of additional factors that have been omitted in this 
one-factor model.

It is important to stress that the observed
increase in conditional correlation between
Disney and the Standard \& Poor's is not due to a genuine increase in the
unconditional correlation but results only from the interplay between the
conditioning and the dependence structure of the factor model.

In summary, the conditional correlation coefficient is an unreliable tool,
to say the least,
for characterizing possible changes in the dependence structure when
the volatility level increases,
as it is often read in the contagion literature.

\subsection{Conditional correlation coefficient on both variables}

\subsubsection{Definition}
We consider two random variables $X$ and $Y$ and define their
conditional correlation coefficient $\rho_{\A,\B}$, conditioned upon
$X \inA$ and $Y \inB$,
where $\A$ and $\B$ are two subsets of $\mathbb R$ such that $\Pr \{X \inA ,
Y \inB \} >0$, by
\be
\label{eq:rho_ab}
\rho_{\A,\B}=\frac{\cov(X, Y ~|~ X \inA, Y \inB)
}{\sqrt{\var(X ~|~ X \inA, Y \inB) \cdot  \var(Y ~|~ X \inA, Y \inB)}}~.
\ee

In this case, it is much more difficult to obtain general results for any
specified class of distributions compared to the previous case of
conditioning on a single
variable. We have only been able to give the asymptotic behavior for a Gaussian
distribution in the situation detailed below, using
the expressions in \cite[p 113]{JK72} or proposition A.1 of \cite{AC01}.

\subsubsection{Conditioning on $X$ and $Y$ larger than $u$ for bivariate
Gaussian random variables}

Let us assume that the pair of random variables (X,Y) has a Normal distribution
with unit unconditional variance and unconditional correlation coefficient
$\rho$. The subsets $\A$ and $\B$ are both choosen equal to $[u, +\infty)$,
with $u \in \mathbb R_+$. Let us denote by $\rho_u$ the correlation coefficient
conditional on this particular choice for the subsets $\A$ and $\B$:
\be
\rho_u=\frac{\cov(X, Y ~|~ (X,Y) \in [u, +\infty)^2)
}{\sqrt{\var(X ~|~(X,Y) \in [u, +\infty)^2) \cdot  \var(Y ~|~ (X,Y) \in [u,
+\infty)^2)}}~.
\ee
Under the assumptions stated above, Appendix
\ref{app:ds_ccc_gv} shows that, for large $u$:
\be
\rho_u \sim_{u \rightarrow \infty} \rho ~ \frac{1+\rho}{1-\rho} \cdot
\frac{1}{u^2}~,
\label{mgjzma}
\ee
which goes to zero. This decay is faster than in the case
governed by (\ref{eq:os_ccc_gv}) resulting from the
conditioning on a single variable leading to
$\rho_v^+ \sim_{v \rightarrow +\infty} ~1/v$, but, as we expect, we do not
observe a significant change. Thus, the correlation coefficient
conditioned on both variables does not yield new information and does not bring
any special improvement with respect to the correlation coefficient conditioned
on a single variable.

%********************************
%*              Section 3                *
%********************************

\section{Conditional concordance measures}

The (conditional) correlation coefficient, which has just been
investigated, suffers
from several deficiencies. First, it is only a measure of linear
dependence. Thus, as stressed by \cite{EMN99}, it is fully satisfying
only for the description of the dependence of variables with elliptical
distributions. Moreover, the correlation coefficient aggregates the
information contained both in the marginal and in the
collective behavior. The correlation coefficient is not invariant 
under an increasing change
of variable, a transformation which is known to let unchanged the dependence
structure. Thus, it is desirable to find another measure
of the dependence between two assets or more generally between two
random variables, which, contrarily to the linear correlation
coefficient, only depends on the copula properties, and are therefore not
affected by a change in the marginal distributions (provided
that the mapping is increasing). It turns out that this
desirable property is shared
by all {\it concordance measures}. Among these measures are the 
well-known Kendall's tau,
Spearman's rho or Gini's beta (see \cite{Nelsen} for details).

However, these concordance
measures are not well-adapted, as such,  to the study of extreme
dependence, because they are functions of the whole distribution, including
the moderate and small returns.
A simple idea to investigate the extreme concordance
properties of two random variables
is to calculate these
quantities conditioned on values larger than a given threshold and let this
threshold go to infinity.

In the sequel, we will only focus on the Spearman's rho which can be
easily estimated empirically. It offers a natural
generalization of the (linear) correlation coefficient. Indeed, the
correlation coefficient quantifies the degres of linear dependence
between two random variables, while the Spearman's rho quantifies the
degree of functional dependence, whatever the functional dependence
between the two random variables may be. This represents a very
interesting improvement. Perfect correlations (resp. anti-correlation)
give a value $1$ (resp. $-1$) both for the standard correlation coefficient
and for the Spearman's rho. Otherwise, there is no general relation allowing
to deduce the Spearman's rho
from the correlation coefficient and vice-versa.

\subsection{Definition}

The Spearman's rho, denoted by $\rho_s$ in the sequel, measures the difference
between the probability of concordance and the probability of discordance for
the two pairs of random variables $(X_1, Y_1)$ and $(X_2, Y_3)$, where
the pairs $(X_1, Y_1)$, $(X_2, Y_2)$ and $(X_3, Y_3)$ are three independent
realizations drawn from the same distribution:
\be
\rho_s = 3 \left( \Pr[(X_1-X_2)(Y_1-Y_3)>0]-\Pr[(X_1-X_2)(Y_1-Y_3)<0]
\right).
\ee

The Spearman's rho can also be expressed with the copula $C$ of the two
variables X and Y (see \cite{Nelsen}, for instance):
\be
\label{eq:rs_def}
\rho_s = 12 \int_0^1 \int_0^1 C(u,v)~du~dv -3,
\ee
which allows us to easily calculate $\rho_s$ when the copula $C$ is known in
closed form.

Denoting $U=F_X(X)$ and $V=F_Y(V)$, it is easy to show that $\rho_s$ is
nothing but the (linear) correlation coefficient of the uniform random
variables $U$ and $V$ : \be
\label{eq:rs_2}
\rho_s = \frac{\cov(U,~V)}{\sqrt{\var(U) \var(V)}}~.
\ee
This justifies its name as a {\it correlation coefficient of the rank},
and shows that it can easily be estimated.

An attractive feature of the Spearman's rho is to be independent of the
margins, as we can see in equation (\ref{eq:rs_def}). Thus, contrarily to the
linear correlation coefficient, which aggregates the marginal properties of the
variables with their collective behavior, the rank correlation 
coefficient takes
into account only the dependence structure of the variables.

Using expression (\ref{eq:rs_2}), we propose a natural definition of the
conditional rank correlation, conditioned on $V$ larger than a given
threshold $\v$:
\be
\rho_s(\v) =  \frac{\cov(U,V~|~V \ge \v)}{\sqrt{\var(U~|~ V \ge \v)
\var(V~|~ \ge \v)}}~.
\label{mmtrlg}
\ee
Appendix \ref{app:sp} shows that $\rho_s(\v)$ can be expressed as a 
functional form
of the copula only:
\be
\label{eq:sp}
\rho_s(\v) =\frac{\frac{12}{1-\v} \int_\v^1 dv \int_0^1 du~C(u,v) -
6 \int_0^1 du
~ C(u,\v) - 3}{\sqrt{
1-4 \v + 24~(1-\v) \int_0^1
du~ u~C(u,\v) + 12~( 2\v-1) \int_0^1 du ~ C(u,\v)
- 12 \left( \int_0^1 du ~ C(u,\v) \right)^2 }}~.
\ee

\subsection{Example}

Contrarily to the conditional correlation coefficient, we have not been able to
obtain analytical expressions for the conditional Spearman's rho,
at least for the distributions that we have considered up to now.  Obviously,
for many families of copulas known in closed form, equation 
(\ref{eq:sp}) allows
for an explicit calculation of $\rho_s(v)$. However, most copulas of 
interest in
finance have no simple closed form, so that it is necessary to resort to
numerical computations.

\begin{figure}
\begin{center}
\includegraphics[width=15cm]{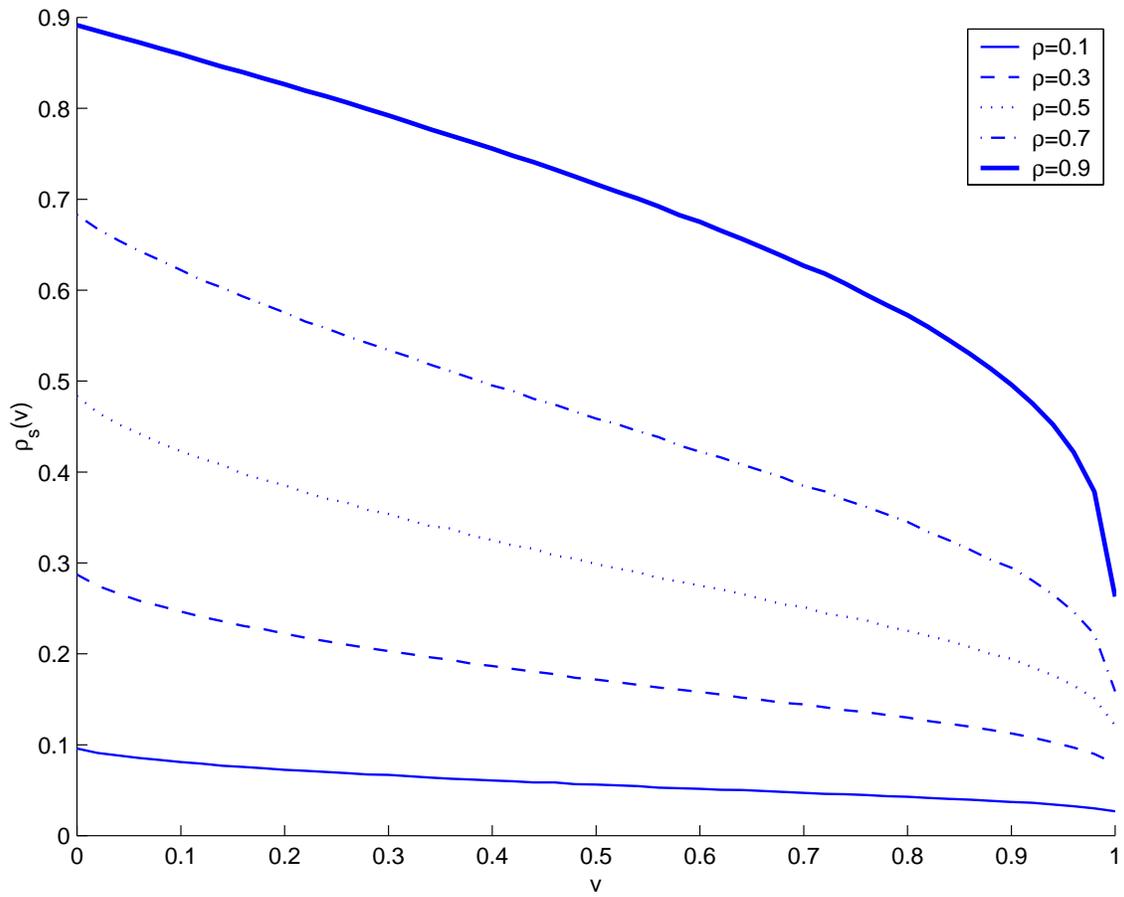}
\caption{\label{fig:spearman} Conditional Spearman's rho for bivatriate
Gaussian distributions with unconditional linear correlation coefficient $\rho
= 0.1, 0.3, 0.5, 0.7, 0.9$. }
\end{center}
\end{figure}

As an example, let us consider the bivariate Gaussian distribution (or copula)
with unconditional correlation coefficent $\rho$. It is well-known that its
unconditional Spearman's rho is given by
\be
\rho_s = \frac{6}{\pi} \cdot \arcsin \frac{\rho}{2} ~ .
\label{rhospear}
\ee
Figure \ref{fig:spearman} shows the conditional Spearman's rho $\rho_s(v)$
defined by (\ref{mmtrlg}) obtained from a numerical
integration of (\ref{eq:sp}). We observe the same bias as for the conditional
correlation coefficient, namely the conditional rank correlation changes with
$v$ eventhough the unconditional correlation is fixed to a constant 
value. Nonetheless, this
conditional Spearman's rho seems more sensitive than the conditional
correlation coefficient since we can observe in figure \ref{fig:spearman} that,
as $v$ goes to one, the conditional Spearman's rho $\rho_s(v)$ does not go to
zero for all  values of $\rho$, as previously observed with the conditional
correlation coefficient (see equation (\ref{eq:os_ccc_gv})).

\subsection{Empirical evidence}

\begin{figure}
\begin{center}
\includegraphics[width=15cm]{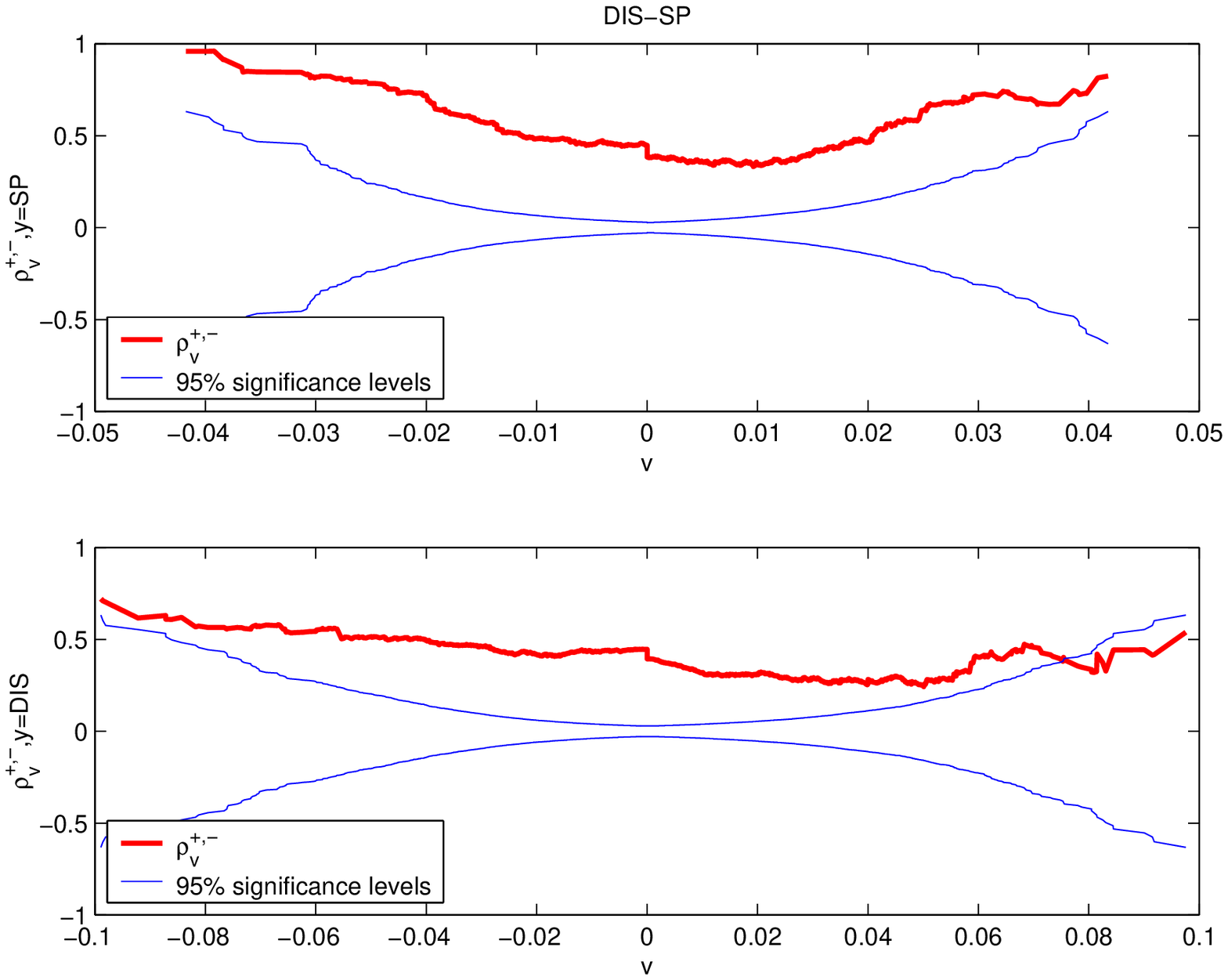}
\caption{\label{fig:corr_dis} In the upper panel, the thick curve depicts
the correlation coefficient between the Disney Company daily returns and the
Standard \& Poor's 500 index daily returns conditional on the Standard \&
Poor's 500 index daily returns larger than (smaller than) a given positive
(negative) value $v$. The two thin curves represent the area within which we
cannot consider, at the 95\% confidence level, that the estimated correlation
coefficient is significantly different from zero. The lower panel gives the
same kind of information but for the correlation coefficient conditioned on the
Disney Company daily returns larger than (smaller than) a given positive
(negative) value $v$.}
\end{center}
\end{figure}

\begin{figure}
\begin{center}
\includegraphics[width=15cm]{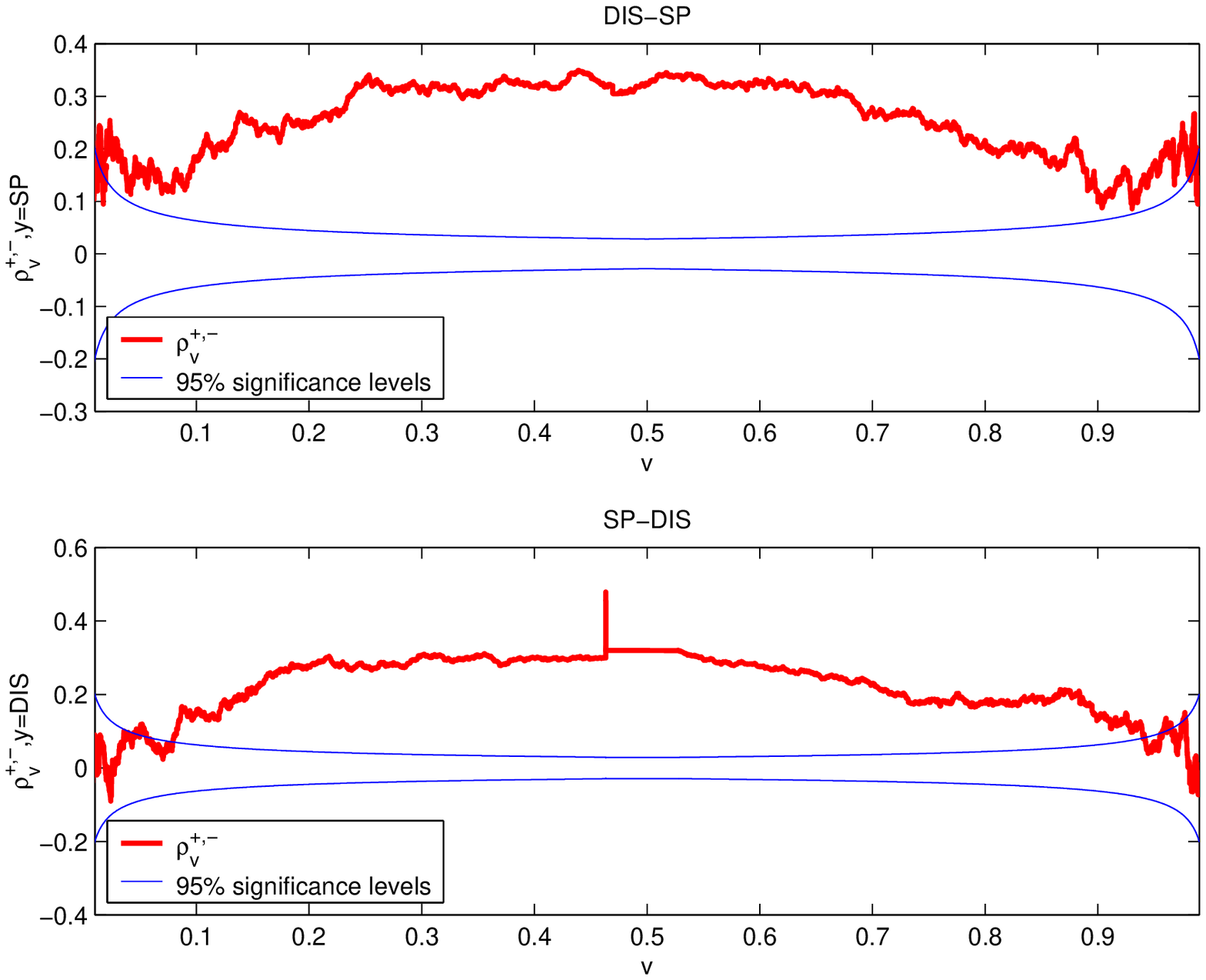}
\caption{\label{fig:spear_dis} In the upper panel, the thick curve depicts
the Spearman's rho between the Disney Company daily returns and the
Standard \& Poor's 500 index daily returns. Above the quantile $v=0.5$, the
Spearman's rho is conditioned on  the Standard \&
Poor's 500 index daily returns whose quantile is larger than v, while
below the quantile $v=0.5$ it
is conditioned on the Standard \& Poor's 500 index daily returns
whose quantile is smaller than v. The two thin curves represent the
area within which we
cannot consider, at the 95\% confidence level, that the estimated
Spearman's rho is significantly different from zero. The lower panel gives the
same kind of information but for the Spearman's rho conditioned on the
realizations of the Disney Company daily returns.}
\end{center}
\end{figure}

\begin{figure}
\begin{center}
\includegraphics[width=15cm]{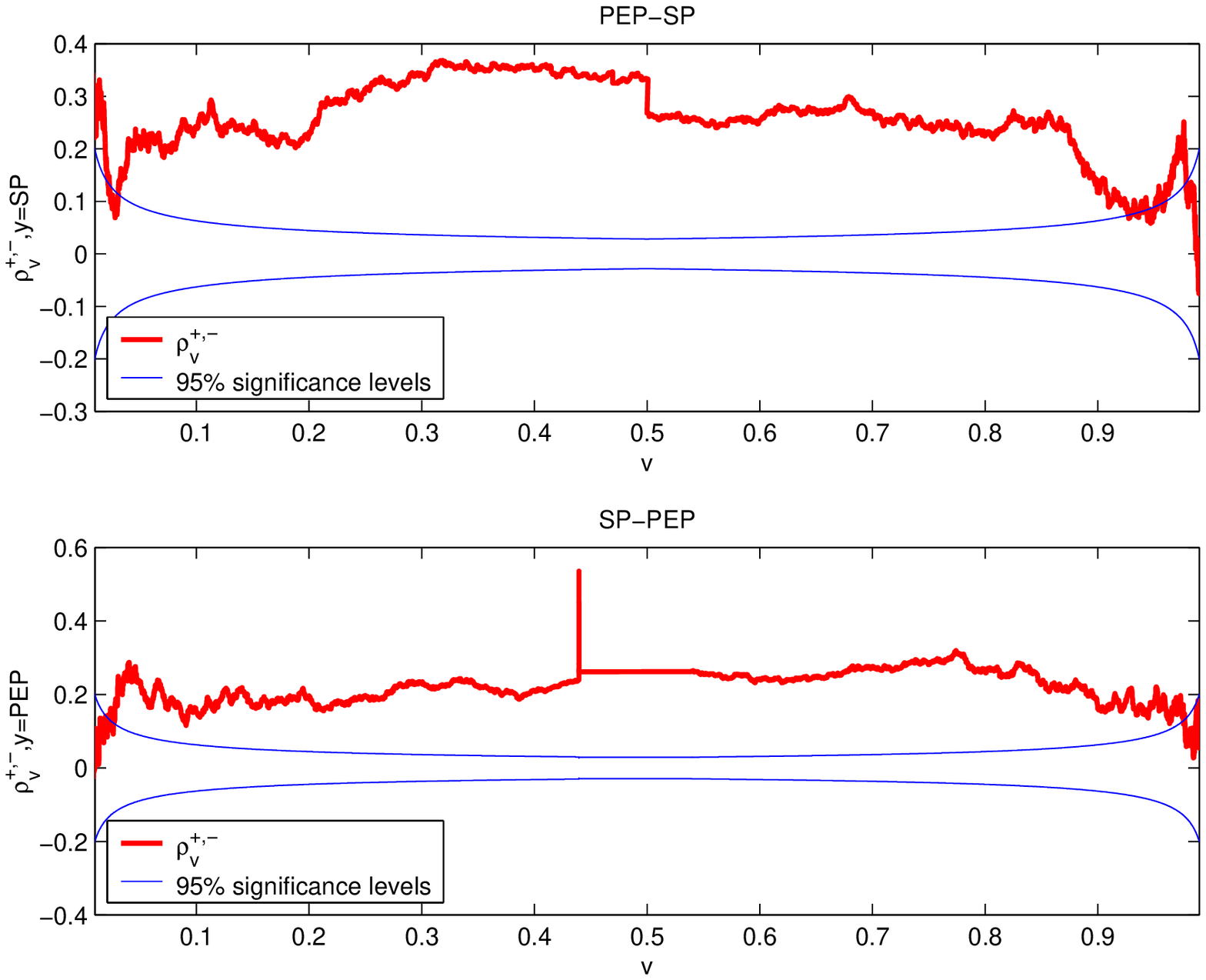}
\caption{\label{fig:spear_pep} In the upper panel, the thick curve depicts
the Spearman's rho between the Pepsico Incorporated daily returns and the
Standard \& Poor's 500 index daily returns. Above the quantile $v=0.5$, the
Spearman's rho is conditioned on  the Standard \&
Poor's 500 index daily returns whose quantile is larger than v, while
below the quantile $v=0.5$ it is conditioned on the Standard \& 
Poor's 500 index
daily returns whose quantile is smaller than v. The two thin curves
represent the area within which we cannot consider, at the 95\%
confidence level, that the estimated Spearman's rho is significantly
different from zero. The lower panel gives the same kind of
information but for the Spearman's rho conditioned on the realizations
of the Pepsico Incorporated daily returns.}
\end{center}
\end{figure}

In figures \ref{fig:spear_dis} and \ref{fig:spear_pep} we have
represented the conditionnal Spearman's rho for the same assets as in
figures \ref{fig:corr_dis} and \ref{fig:corr_pep}. The results are
strikingly different compared to figures (\ref{fig:corr_dis}) and
(\ref{fig:corr_pep}). While the conditional correlation
coefficient would lead to conclude positively on the existence of significant
extreme correlations between Disney and the Standard \& Poor's 500 index,
the conditional Spearman's rho leads to the opposite
conclusion. Specifically, figure
\ref{fig:corr_dis} suggests that the correlation coefficient conditioned on
large negative Standard \& Poor's returns converges to a value very 
close to $1$.
Since Spearman's rho (conditioned or not) goes to $1$ when the 
correlation coefficient
(conditioned or not) goes to $1$, this predicts that the
Spearman's rho conditioned on large negative Standard \& Poor's returns
should also converge to $1$. We observe the opposite behavior that
the Spearman's rho is not
significantly different from zero in the lowest quantiles.

This contradictory empirical result puts in light the problem of obtaining
reliable and sensitive estimations of such correlation measures.
In particular, the Pearson's
coefficient usually employed to estimate the correlation coefficient
between two variables is known to be not very efficient when the
variables are fat-tailed and when the estimation performed on a small
sample. Indeed, with small samples, the Pearson's coefficient is very
sensitive to the largest value, which can lead to an important bias in
the estimation.

Moreover, even with large sample size, \cite{Meerschaert} have shown that the
nature of the convergence between the
Pearson's coefficient of two times series with tail index $\mu$
and the theoretical correlation as the sample size $T$ tends to infinity
is sensitive to the existence and strength of the theoretical correlation.
If there is no theoretical correlation between the two times series, the
sample correlation tends to zero with Gaussian
fluctuations. If the theoretical correlation is non-zero, the
difference between
the sample correlation and the theoretical correlation times
$T^{1-2/\mu}$ converges in distribution to a stable law with index $\mu/2$.

%********************************
%*          Section 4           *
%********************************

\section{Tail dependence}

For the sake of completeness, and since it is directly related to the
multivariate extreme values theory, we study the so-called coefficient of
tail dependence $\lambda$. To our knowledge, its interest for 
financial applications has
been first underlined by \cite{EMS01}.

The coefficient of tail dependence characterizes an important property of the
extreme dependence between $X$ and $Y$, using the (original
or unconditional) copula of $X$ and $Y$. In constrast,
the conditional spearman's rho is defined in terms of a conditional copula, as
it can be seen as the  ``unconditional Spearman's rho'' of the
copula of $X$ and $Y$ conditioned on $Y$ larger than the threshold $v$.
This
copula of $X$ and $Y$ conditioned on $Y$ larger than the threshold $v$
is not the true copula of $X$ and $Y$ because it is modified by the 
conditioning.
In this sense, the tail dependence parameter $\lambda$ is a more 
natural property
directly related to the copula of $X$ and $Y$.

To begin with, we recall the definition of the
coefficient $\lambda$ as well as of $\bar \lambda$ (see below) which 
allows one to quantify the
amount of dependence in the tail. Then, we present several results
concerning the coefficient $\lambda$ of tail dependence for various 
distributions and
models, and finally, we discuss the problems encountered in the estimation of
these quantities.

\subsection{Definition}
The concept of tail dependence is appealing by its simplicity. By
definition,  the
(upper) tail dependence coefficient is:
\be
\lambda=\lim_{u \rightarrow 1} \Pr[ X > F_X^{-1}(u) | Y > F_Y^{-1}(u)]~ ,
\label{ngallqle}
\ee
and quantifies the probability to observe a large $X$, assuming that
$Y$ is large
itself. For a survey of the properties of the tail dependence coefficient, the
reader is refered to \cite{Coles_etal,EMS01,L99}, for instance.
In words, given that $Y$ is very large (which occurs with probability
$1-u$), the probability that $X$ is very large at the same probability
level $u$ defines asymptotically the tail dependence coefficient $\lambda$.

One of the appeal of this definition of tail dependence is that it
is a pure copula property, i.e., it is independent of the
margins of $X$ and $Y$. Indeed, let $C$ be the copula of the
variables $X$ and $Y$, then if the bivariate copula $C$ is such that
\be
\lim_{u \rightarrow 1} \frac{1-2u+C(u,u)}{1-u} = \lim_{u \rightarrow
   1} 2 - \frac{\log C(u,u)}{\log u} =\lambda
\label{nfmsla}
\ee
exists, then $C$ has an upper tail dependence coefficient $\lambda$
(see \cite{Coles_etal,EMS01,L99}).

If $\lambda > 0$, the copula presents tail dependence and large events tend to
occur simultaneously, with the probability $\lambda$. On the contrary,
when $\lambda=0$, the copula has no tail dependence, in this sense,
and the variables $X$ and $Y$ are said asymptotically independent.
There is however a subtlety in this definition (\ref{ngallqle}) of tail
dependence. To make it clear, first consider the case where for large
$X$ and $Y$
the cumulative distribution function $H(x,y)$ factorizes such that
\be
\label{eq:tl}
\lim_{x,y \rightarrow \infty} \frac{F(x,y)}{F_X(x) F_Y(y)}=1~,
\ee
where $F_X(x)$ and $F_Y(y)$ are the margins of $X$ and $Y$ respectively.
This means that, for $X$ and $Y$ sufficiently large, these two
variables can be considered as independent. It is then easy to show that
\bea
\lim_{u \rightarrow 1} \Pr\{ X > {F_X}^{-1}(u) | Y > {F_Y}^{-1}(u) \}&=&
\lim_{u \rightarrow 1} 1-F_X( {F_X}^{-1}(u)) \\
&=&\lim_{u \rightarrow 1} 1-u =0,
\eea
so that independent variables really have no tail dependence $\lambda
=0$, as one
can expect.

However, the result $\lambda = 0$ does not imply that the multivariate
distribution can be automatically factorized asymptotically, as shown 
by the Gaussian
example. Indeed, the Gaussian multivariate distribution
does not have a factorizable multivariate distribution, even 
asymptotically for extreme values,
since the non-diagonal term of the quadratic form in the
exponential function does not become negligible in general as $X$ and
$Y$ go to infinity. Therefore, in a weaker sense, there
may still be a dependence in the tail even when $\lambda = 0$.

To make this statement more precise, following \cite{Coles_etal}, let
us introduce the coefficient
\bea
\bar \lambda &=& \lim_{u \rightarrow 1} \frac{2 \log \Pr\{ X >
{F_X}^{-1}(u)\}}{\log \Pr\{ X > {F_X}^{-1}(u),  Y > {F_Y}^{-1}(u)\}}
-1\\&=& \lim_{u \rightarrow 1} \frac{2 \log (1-u)}{ \log [1-2u +
   C(u,u)]} -1~.
\eea
It can be shown that the coefficient $\bar \lambda = 1$ if and only if
the coefficient of tail dependence $\lambda > 0$, while $\bar \lambda$
takes values in $[-1,1)$ when $\lambda=0$, allowing us to
quantify the strength of the dependence in the tail in such a case.
In fact, it has been established that, when $\bar \lambda >0$, the
variables $X$ and $Y$ are simultaneously large more frequently than
independent variables, while simultaneous large deviations of $X$ and
$Y$ occur less frequenlty  than under independence when $\bar \lambda
<0$ (the interested reader is refered to \cite{LT96,LT98}).

To summarize, independence (factorization of the bivariate distribution)
implies no tail dependence $\lambda =0$. But $\lambda =0$ is not
sufficient to imply factorization and thus true independence. It
also requires, as a necessary condition, that $\bar \lambda=0$.

We will first recall the expression of the tail dependence coefficient for
usual distributions, and then calculate it in the case of a one-factor model
for different distributions of the factor.

\subsection{Tail dependence for Gaussian distributions and Student's
distributions}
\label{sec:ts}

Assuming that $(X,Y)$ are normally distributed with correlation
coefficient $\rho$, \cite{EMS01} shows that for all $\rho \in [-1, 1)$,
$\lambda =0$, while \cite{H00} gives $\bar \lambda = \rho$, which
expresses, as one can expect, that extremes appear more likely
together for positively correlated variables.

In constrast, if $(X,Y)$ have a Student's distribution,
\cite{EMS01} shows that the tail dependence coefficient is
\be
\label{eq:student}
\lambda=2 \cdot \bar T_{\nu+1} \left(\sqrt{\nu+1} \sqrt{\frac{1-\rho}{1+\rho}}
\right),
\ee
which is greater than zero for all $\rho>-1$, and thus $\bar \lambda =
1$. This last example prooves that extremes appear more likely together
whatever the correlation coefficient may be, showing that, in fact, it
does not exist any general relation between the asymptotic dependence
and the linear correlation coeffient.

Let us add that these two distributions are elliptical distributions,
and that the coefficients of tail dependence obtained in the two
previous cases are characteristic of the coefficient of tail dependence
observed for this class of distributions. Indeed, \cite{HL01} have
shown that ellipticaly distributed random variables presents tail
dependence if and only if they are regularly varing, i.e behaves
asymptotically like power laws with exponent $\nu$. In such a case,
for every regularly varying pair of random variables elliptically
distributed, we have
\be
\label{eq:elliptical}
\lambda = \frac{\int_{\pi/2 - \arcsin \rho}^{\pi/2} dt ~ \cos^\nu
   t}{\int_0^{\pi/2} dt ~ \cos^\nu t}.
\ee

Note that necessarily, the rhs of equations (\ref{eq:student}) and
(\ref{eq:elliptical}) are equal, which is natural since the
correlation coefficient is an invariant quantity in the class
of elliptical distributions and that the coefficient of tail
dependence is only determined by the asymptotic behavior of the
distribution, so that it does not matter that the distribution is a
Student's distribution with $\nu$ degrees of freedom or any other
elliptical distribution with the same behavior in the tail.

\subsection{Tail dependence generated by a factor model}

Consider the one-factor model
\be
X=\alpha Y+ \epsilon,
\ee
where $\epsilon$ is a random variable independent of $Y$ and $\alpha$
a non-random
{\it positive} coefficient.
We now study the tail dependence coefficient $\lambda$
between the two random variables $X$ and $Y$ in two simple cases.

\subsubsection{The Gaussian case}
This case is trivial. Indeed, assume that the factor $Y$ and the idiosyncratic
noise $\epsilon$ have Gaussian distributions, it is then obvious that the
bivariate distribution of $(X,Y)$ is Gaussian. Thus, the tail dependence
coefficient generated by a one factor model, with a Gaussian factor equals
zero.

In fact, as we shall see in the sequel, this result remains true
for all distributions of the idiosyncratic noise $\epsilon$.

\subsubsection{The Student's case}
Let us assume now that the factor $Y$ and the idiosyncratic noise
$\epsilon$ have
centered Student's distributions with the same number $\nu$ of
degrees of freedom and scale
factors respectively equal to $1$ and $\sigma$. The choice of the
scale factor equal to
$1$ for $Y$ is not restrictive but only provides a convenient normalization for
$\sigma$. Appendix \ref{app:tls}
shows that the tail dependence coefficient is
\be
\lambda = \frac{1}{1+ \left( \frac{\sigma}{\alpha} \right) ^\nu}~.
\label{mhjkkn}
\ee
As is reasonable intuitively, the larger the typical scale $\sigma$
of the fluctuation
of $\epsilon$ and the weaker is the coupling coefficient $\alpha$,
the smaller is the tail dependence.

Let us recall that the unconditional correlation coefficient
$\rho$ can be writen as $\rho= (1 + \frac{ \sigma^2}{ \alpha^2} )^{-1/2} $,
which allows us to rewrite the   coefficient of upper tail dependence as
\be
\lambda= \frac{\rho^\nu}{ \rho^\nu + (1- \rho^2)^{\nu/2}} ~.
\ee

Surprinsingly, $\lambda$ does not go to zero for all
$\rho$'s as $\nu$ goes to infinity, as one would expect intuitively.
Indeed, a natural reasoning would be that,
as $\nu$ goes to infinity, the Student's
distribution goes to the Gaussian distribution. Therefore, one could
{\it a priori} expect to find again the result given in the previous 
section for the
Gaussian factor model. We note that $\lambda \to 0$ when $\nu \to \infty$
for all $\rho$'s smaller than $1/\sqrt{2}$. But, and here lies the
surprise, $\lambda \to 1$ for all $\rho$
larger than $1/\sqrt{2}$ when $\nu \to \infty$. This
counter-intuitive result is due to a non-uniform convergence which
makes the order to two limits non-commutative: taking first
the limit $u \to 1$ and then $\nu \to \infty$ is different from
taking first the limit $\nu \to \infty$ and then $u \to 1$.
In a sense, by taking first the limit $u \to 1$, we always ensure
somehow the power law regime even if $\nu$ is later taken to 
infinity. This is different
from first ``sitting'' on the Gaussian limit $\nu \to \infty$. It then is
a posteriori reasonable that the absence of uniform convergence is made
strongly apparent in its consequences when measuring a quantity
probing the extreme tails of the distributions.

\begin{figure}
\begin{center}
\includegraphics[width=8cm]{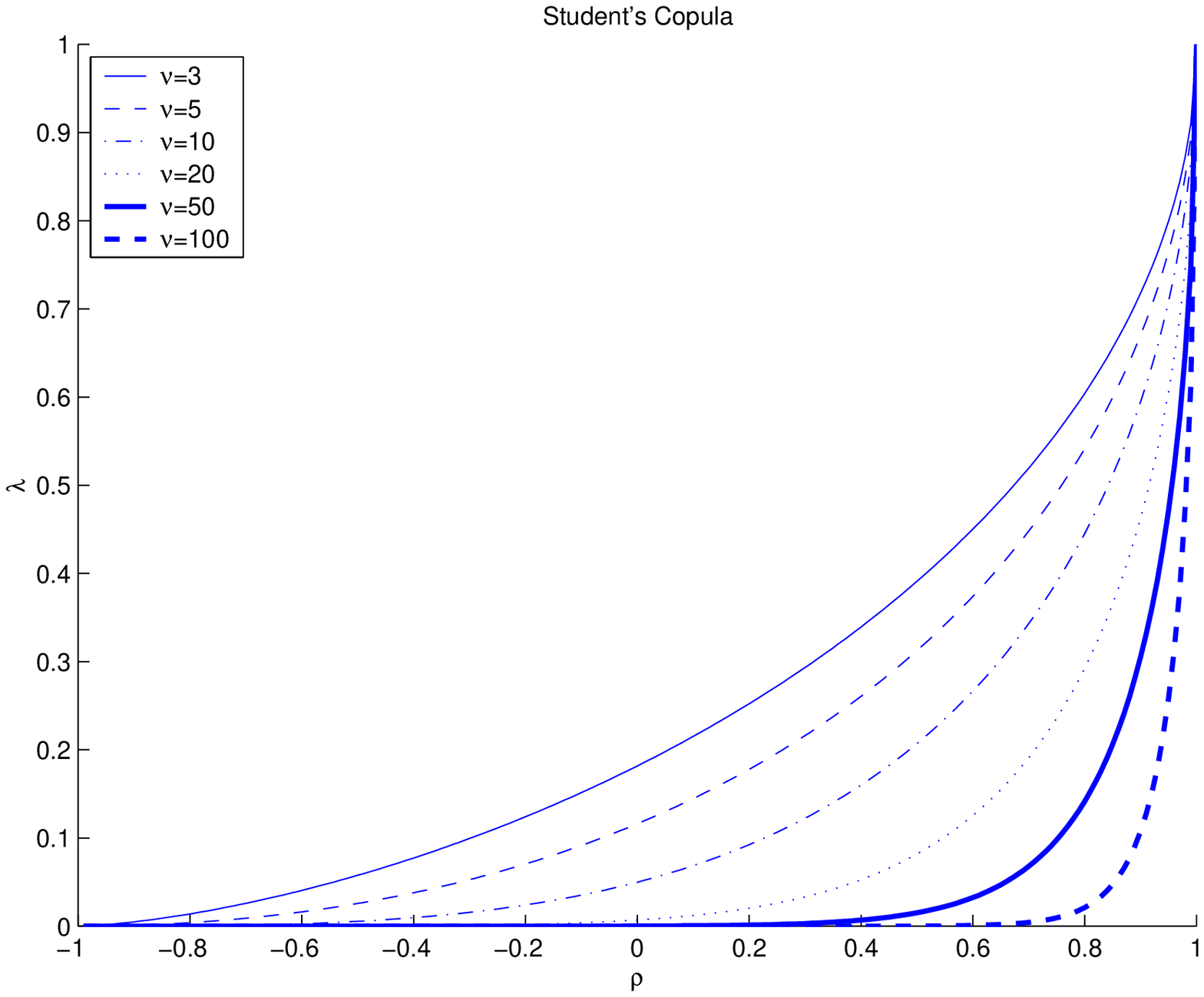}
\includegraphics[width=8cm]{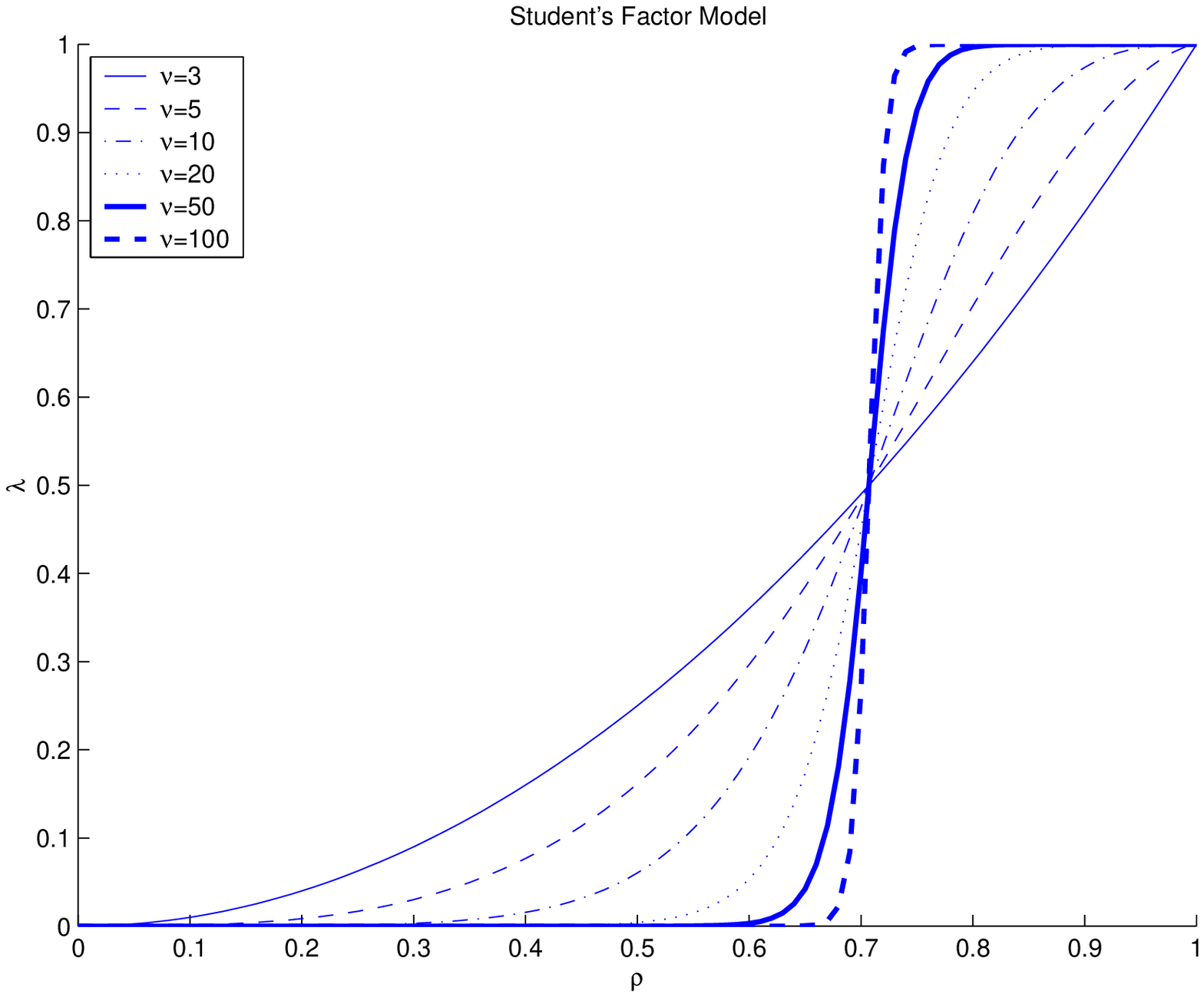}
\caption{\label{fig:lambda} Coefficient of upper tail dependence as a
   function of the correlation coefficient $\rho$ for various values of
   the number of degres of freedomn $\nu$ for the student's Copula (left
   panel) and the Student's factor model (right panel).}
\end{center}
\end{figure}

As an illustration, figure \ref{fig:lambda} represents the 
coefficient of tail dependence for the
Student's copula and Student's factor model as a function of $\rho$ for
various value of $\nu$. It is interesting to note that $\lambda$ equals zero
for all negative $\rho$ in the case of the factor model, while
$\lambda$ remains non-zero for negative values of the correlation coefficient
for bivariate Student's variables.

If $Y$ and $\epsilon$ have different numbers $\nu_Y$ and $\nu_{\epsilon}$
of degrees of freedom, two cases occur. For $\nu_Y < \nu_{\epsilon}$,
$\epsilon$
is negligible asymptotically and $\lambda =1$. For $\nu_Y > \nu_{\epsilon}$,
$X$ becomes identical to $\epsilon$ because $Y$ is negligible
asymptotically. Thus $X$ and $Y$ have the same tail-dependence as
$\epsilon$ and $Y$,
which is $0$ by construction.

\subsubsection{General case}

A general result concerning the tail dependence generated by factor
models for every kind of factor and noise distributions has recently been
established by \cite{MS02}. It has been proved that the coefficient
of (upper) tail dependence between $X$ and $Y$ is given by
\be
\lambda = \int_{\max \left\{ 1, \frac{l}{\alpha} \right\} }^\infty dx ~f(x)~,
\label{mhjjskasz}
\ee
where, provided that they exist,
\bea
l &=& \lim_{u \rightarrow 1} \frac{{F_X}^{-1}(u)}{{F_Y}^{-1}(u)},\\
f(x) &=& \lim_{t \rightarrow \infty}\frac{t \cdot P_Y(t \cdot x)}{\bar
   F_Y(t)}~.
\eea
In the framework of this general theorem (\ref{mhjjskasz}), the previous
result (\ref{mhjkkn}) is a special case which can be easily retrieved.

As a direct consequence, one can show that any rapidly varying factor
leads to a vanishing coefficient of tail dependence, whatever the
distribution of the idiosyncratic noise may be. On the contrary,
regularly vaying factors lead to a tail dependence, provided that the
distribution of the idiosycratic noise does not become fatter-tailed than
the factor distribution. One can thus conclude that, in order to
generate tail dependence, the factor must have a sufficiently `wild'
distribution.

\subsection{Estimation of the coefficient of tail dependence}

It would seem that the coefficient of tail dependence could provide a
useful measure of the extreme dependence between two random
variables. Indeed, it could be an interesting quantity to study for
risk management purposes since it gives directly the probabilty that an
asset suffers a large loss assuming that a large loss occured for
another asset. It is also important for contagion
problems. Indeed, the evolution of the coefficient of tail dependence
with the volatility of the markets could be a very significant
indicator of possible contagions between these markets.

Unfortunately, the empirical estimation of the coefficient of tail
dependence is a strenuous task, and to our knowledge, there is not yet any
reliable estimator.
A direct estimation of the conditional probability $\Pr \{ X >
{F_X}^{-1}(u) ~|~ Y > {F_Y}^{-1}(u) \}$, which should tend to $\lambda$ when
$u \to 1$ is impossible to put in practice due to the combination of
the curse of
dimensionality and the drastic decrease of the number of
realisations as $u$ become close to one.

A better approach consists in using kernel estimators, which generally
provide smooth and accurate estimators \cite{K99, Li_etal,S00}. However, these
smooth estimators lead to differentiable estimated copulas which have
automatically vanishing tail dependence.
Indeed, in order to obtain a non-vanishing coefficient
of tail dependence, it is necessary for the corresponding copula to be
non-differentiable at the point $(1,1)$ (or at $(0,0)$).

An alternative is the parametric approach. One can choose to model
dependence via a specific copula, and thus to determine the associated
tail dependence \cite{LS01, MS01, P01}. The
problem of such a method is that the choice of the parameterization
of the copula amounts to choose a priori whether or not the data
presents tail dependence.

An improvement in the estimation of tail dependence has been obtained recently
by using the semi-parametric approach of \cite{MS02}. This approach is based
on the existence of a representation of dependence in terms of a factor
model. In this approach, the estimation of the tail dependence relies 
solely on the
estimation of the marginal distributions, a significantly easier task.

%******************************
%*        conclusion           *
%******************************

\section{Summary and Discussion}

Table \ref{table1} summarizes the asymptotic dependences for large
$v$ and $u$ of
the signed conditional correlation
coefficient $\rho_{v}^+$, the unsigned conditional correlation
coefficient $\rho_{v}^s$ and
the correlation coefficient $\rho_{u}$ conditioned on both variables
for the bivariate
Gaussian, the Student's model, the Gaussian factor model and the
Student's factor model.
Our results provide a quantitative proof that conditioning on
exceedance leads to
conditional correlation coefficients that may be very different from
the unconditional
correlation. This provides a straightforward mechanism for
fluctuations or changes of
correlations, based on fluctuations of volatility or changes of
trends. In other words,
the many reported variations of correlation structure may be in large
part attributed
to changes in volatility. For instance,
our result (\ref{mghhek}) for the Student's factor model
explains the Monte-Carlo and empirical results \cite{Cizeau_etal} that
the unsigned conditional correlation
$\rho_{v}^s$ increases with the exceedance level $v$.

We also suggest that the distinct dependences as a function
of exceedance $v$ and $u$ of the conditional
correlation coefficients may offer novel tools for characterizing
the statistical multivariate distributions of extreme events.
Since their direct characterization is in general restricted by the
curse of dimensionality
and the scarsity of data,
the conditional correlation coefficients provide reduced robust
statistics which can be
estimated with reasonable accuracy and reliability.

Table \ref{table2} gives the asymptotic values of
$\rho_{v}^+$, $\rho_{v}^s$ and $\rho_{u}$ for $v \to +\infty$ and $u
\to \infty$
in order to compare them with the tail-dependence $\lambda$.

These two tables only scratch the surface of the rich sets of measures of tail
and extreme dependences. We have shown that complete
independence (factorization of the bivariate distribution)
implies no tail dependence $\lambda =0$. But $\lambda =0$ does not
imply factorization.
Conversely, tail dependence $\lambda >0$ implies
absence of factorization. But $\lambda >0$ does not imply necessarily that
the conditional correlation coefficients $\rho_{v=\infty}^+$ and
$\rho_{v=\infty}^s$
are non-zero.

Note that the examples of Table \ref{table2} are such that
$\lambda = 0$ seems to go hand-in-hand with $\rho_{v \to \infty}^+ =
0$. However, the
logical implication $(\lambda = 0) ~~ \Rightarrow ~~ (\rho_{v \to
\infty}^+ = 0)$
does not hold in general. A counter example is offered
by the Student's factor model in the case where
$\nu_Y > \nu_{\epsilon}$ (the tail of the distribution of the
idiosynchratic noise is fatter than that
of the distribution of the factor). In this case,
$X$ and $Y$ have the same tail-dependence as $\epsilon$ and $Y$, which is $0$
by construction. But,  $\rho_{v=\infty}^+$ and $\rho_{v=\infty}^s$ are both
$1$ because a large $Y$ almost always gives a large $X$ and the simultaneous
occurrence of a large $Y$ and a large $\epsilon$ can be neglected.
The reason for this absence of tail dependence (in the sense of $\lambda$)
coming together with asymptotically strong conditional correlation coefficients
stems from two facts:
\begin{itemize}
\item first, the conditional correlation
coefficients put much less weight on the extreme tails that the tail-dependence
parameter $\lambda$. In other words, $\rho_{v=\infty}^+$
and $\rho_{v=\infty}^s$ are sensitive to the marginals, i.e.,
there are determined by the full bivariate distribution, while, as we said,
$\lambda$ is a pure copula property independent of the marginals. Since
$\rho_{v=\infty}^+$ and $\rho_{v=\infty}^s$ are measures of tail dependence
weighted by the specific shapes of the marginals, it is natural that they may
behave differently.
\item Secondly, the tail dependence $\lambda$ probes the extreme dependence
property of the original copula of the random variables $X$ and $Y$. On the
contrary, when conditioning on $Y$, one changes the copula of $X$ and $Y$, so
that the extreme dependence properties investigated by the conditional
correlations are not exactly those of the original copula. This last remark
explains clearly why we observe what \cite{Boyer_etal} call a ``bias'' in the
conditional correlations. Indeed, changing the dependence between two random
variables obviously leads to change their correlations.
\end{itemize}

The consequences are potentially of importance. In such a situation, one
measure ($\lambda$) would conclude on
asymptotic tail-independence while the other
measures $\rho_{v=\infty}^+$ and $\rho_{v=\infty}^s$ would conclude
the opposite. Thus, before concluding on a change in the dependence
structure with respect to a given parameter - the volatility or the trend, for
instance - one should check that this change does not result from the tool
used to probe the dependence. In this respect, as recently stressed by
\cite{FR01}, many previous contagion studies may be unreliable.

These questions related to state-varying-dependence 
are also very important for practical applications. As stressed by \cite{AB00, AC01} for
instance, the optimal portfolio will also become state-dependent, and
neglecting this point can lead to very inefficient asset allocation. We would
like to add that, for this purpose,  the time scales involved - i.e., the
investment horizon -will determine which measure is most appropriate.

%********************************
%*                 Appendix               *
%********************************

\newpage
\appendix

%****************************
%*                Appendix 1             *
%****************************

\section{Conditional correlation coefficient for Gaussian variables}
Let us consider a pair of Normal random variables $(X, Y) \sim {\cal N}(0,
\Sigma)$ where $\Sigma$ is their covariance matrix with unconditional
correlation
coefficient $\rho$. Without loss of generality, and for simplicity, we shall
assume $\Sigma$ with unit unconditional variances.

\subsection{Conditioning on $Y$ larger than $v$}
\label{app:os_ccc_gv}

Given a conditioning set $\A = [v, +\infty)$, $v \in {\mathbb R_+}$,
$\rho_\A = \rho_v^+$ is the correlation coefficient conditioned on $Y$ larger
than $v$:
\be
\rho_v^+=
\frac{\rho}{\sqrt{\rho^2+\frac{1-\rho^2}{\var(Y~|~Y > v)}}}.
\ee

We start with the calculation of the first and the second moment of
$Y$ conditioned on $Y$ larger than $v$:
\bea
\E(Y~|~Y>v) &=&\frac{\sqrt{2}}{\sqrt{\pi} e^{\frac{v^2}{2}} \mbox{erfc} \left(
\frac{v}{\sqrt{2}} \right)}= v+\frac{1}{v}-\frac{2}{v^3} + {\cal O} \left(
\frac{1}{v^5} \right), \\
\E(Y^2~|~Y>v) &=&1+ \frac{\sqrt{2}v}{\sqrt{\pi}
e^{\frac{v^2}{2}} \mbox{erfc} \left( \frac{v}{\sqrt{2}} \right)} =
v^2+2-\frac{2}{v^2} + {\cal O} \left( \frac{1}{v^4} \right),  \label{jghhdj}
\eea
which allows us to obtain the variance of $Y$ conditioned on $Y$
larger than $v$:
\be
\var(Y~|~Y > v) =  1+ \frac{\sqrt{2}v}{\sqrt{\pi}
e^{\frac{v^2}{2}} \mbox{erfc} \left( \frac{v}{\sqrt{2}} \right)} - \left(
\frac{\sqrt{2}}{\sqrt{\pi} e^{\frac{v^2}{2}} \mbox{erfc} \left(
\frac{v}{\sqrt{2}} \right)} \right)^2
=  \frac{1}{v^2} +{\cal O} \left( \frac{1}{v^4} \right),   \label{nygwbbs}
\ee
which, for large $v$, yields:
\be
\rho_v^+ \sim_{v \rightarrow \infty} \frac{\rho}{\sqrt{1-\rho^2}}\cdot
\frac{1}{v}~.
\ee

\subsection{Conditioning on $|Y|$ larger than $v$}
\label{app:os_ccc_gv2}

Given a conditioning set $\A =(- \infty, -v] \cup [v, +\infty)$, $v \in
{\mathbb R_+}$, $\rho_\A = \rho_v^s$ is the correlation coefficient
conditioned on
$|Y|$ larger than $v$:
\be
\rho_v^s=
\frac{\rho}{\sqrt{\rho^2+\frac{1-\rho^2}{\var(Y~|~|Y| > v)}}}.
\ee

The first and second moment of $Y$ conditioned on $|Y|$ larger than $v$ can be
easily calculated:
\bea
\E(Y~|~|Y|>v) &=&0, \\
\E(Y^2~|~|Y|>v) &=&1+ \frac{\sqrt{2}v}{\sqrt{\pi}
e^{\frac{v^2}{2}} \mbox{erfc} \left( \frac{v}{\sqrt{2}} \right)} =
v^2+2-\frac{2}{v^2} + {\cal O} \left( \frac{1}{v^4} \right)~.  \label{mgslw}
\eea
Expression (\ref{mgslw}) is the same as (\ref{nygwbbs}) as it should.
This gives the following conditional variance:
\be
\var(Y~|~|Y| > v) =  1+ \frac{\sqrt{2}v}{\sqrt{\pi}
e^{\frac{v^2}{2}} \mbox{erfc} \left( \frac{v}{\sqrt{2}} \right)} =  v^2+ 2
+{\cal O} \left( \frac{1}{v^2} \right)~,
\ee
and finally yields, for large $v$,
\be
\rho_v^s \sim_{v \rightarrow \infty}
\frac{\rho}{\sqrt{\rho^2+\frac{1-\rho^2}{2+v^2}}} \sim_{v \rightarrow \infty}
1- {1 \over 2}~{1-\rho^2 \over \rho^2}~{1 \over v^2}~.
\ee

\subsection{Conditioning on both $X$ and $Y$ larger than $u$}
\label{app:ds_ccc_gv}

By definition, the conditional correlation
coefficient $\rho_u$, conditioned on both $X$ and $Y$ larger than $u$, is
\bea
\rho_u &=& \frac{\cov [X,Y~|~ X>u, Y>u]}{\sqrt{\var[X~|~ X>u, Y>u]}
\sqrt{\var[Y~|~ X>u, Y>u]}}~,\\
&=& \frac{m_{11}-m_{10}\cdot m_{01}}{\sqrt{m_{20}-{m_{10}}^2}
\sqrt{m_{02}-{m_{01}}^2}}~,  \label{ngkkekems}
\eea
where $m_{ij}$ denotes $\E[X^i \cdot Y^j ~|~X>u, Y>u$].

Using the proposition A.1 of \cite{AC01} or the expressions in \cite[p
113]{JK72}, we can assert that
{\small
\bea
m_{10}~L(u,u;\rho) &=& (1+ \rho)~ \phi(u) \left[ 1- \Phi
\left(\sqrt{\frac{1-\rho}{1+\rho}} u \right) \right],   \label{mfdd}\\
m_ {20}~L(u,u;\rho) &=& (1+\rho^2)~ u~ \phi(u) \left[1-
\Phi\left(\sqrt{\frac{1-\rho}{1+\rho}} u \right) \right] +
\frac{\rho~\sqrt{1-\rho^2}}{\sqrt{2 \pi}}~\phi \left( \sqrt{
\frac{2}{1+\rho}} u \right) + L(u,u; \rho),  \label{njhjjnnsw}\\
m_{11}~L(u,u;\rho)&=&2\rho~u~\phi(u) \left[1- \Phi \left(
\sqrt{\frac{1-\rho}{1+\rho}} u \right) \right]
+\frac{\sqrt{1-\rho^2}}{\sqrt{2 \pi}}~\phi \left( \sqrt{ \frac{2}{1+\rho}} u
\right) + \rho~L(u,u;\rho)~,   \label{mmjhmjmr}
\eea}
where $L(\cdot , \cdot ; \cdot)$ denotes the bivariate Gaussian survival (or
complementary cumulative)
distribution:
\be
L(h,k;\rho) = \frac{1}{2 \pi \sqrt{1-\rho^2}} \int_h^\infty dx \int_k^\infty
dy~ \exp \left(-\frac{1}{2} \frac{x^2-2 \rho x y +y^2}{1-\rho^2} \right)~,
\ee
$\phi(\cdot)$ is the Gaussian density:
\be
\phi(x) = \frac{1}{\sqrt{2 \pi}}~e^{-\frac{x^2}{2}},
\ee
and $\Phi(\cdot)$ is the cumulative Gaussian distribution:
\be
\Phi(x) = \int_{-\infty} ^x du~ \phi(u).
\ee

\subsubsection{Asymptotic behavior of $L(u,u;\rho)$}

We focus on the asymptotic behavior of
\be
L(u,u;\rho) = \frac{1}{2 \pi \sqrt{1-\rho^2}} \int_u^\infty dx \int_u^\infty
dy~ \exp \left(-\frac{1}{2} \frac{x^2-2 \rho x y +y^2}{1-\rho^2} \right),
\ee
for large $u$. Performing the change of variables $x'= x-u$ and $y'= y-u$, we
can write
\be
\label{eq:Luu}
L(u,u;\rho) = \frac{e^{-\frac{u^2}{1+\rho}}}{2 \pi \sqrt{1-\rho^2}}
\int_0^\infty dx' \int_0^\infty dy'~ \exp \left(-u \frac{x'+y'}{1+\rho}\right)~
\exp \left(-\frac{1}{2} \frac{x'^2-2 \rho x' y' +y'^2}{1-\rho^2} \right). \ee
Using the fact that
\be
\label{eq:exp}
\exp \left(-\frac{1}{2} \frac{x'^2-2 \rho x' y' +y'^2}{1-\rho^2} \right) = 1-
\frac{x'^2-2\rho x'y'+y'^2}{2(1-\rho^2)} +
\frac{(x'^2-2\rho x'y'+y'^2)^2}{8(1-\rho^2)^2} - \frac{(x'^2-2\rho
x'y'+y'^2)^3}{48(1-\rho^2)^3} + \cdots,
\ee
and applying theorem 3.1.1 in \cite[p 58]{J95} (Laplace's method), equations
(\ref{eq:Luu}) and (\ref{eq:exp}) yield
\bea
\label{eq:Lu_as}
L(u,u;\rho) =  \frac{(1+\rho)^2}{2 \pi \sqrt{1-\rho^2}}
\cdot \frac{e^{-\frac{u^2}{1+\rho}}}{u^2} \left[ 1- \frac{(2-\rho)(1+
\rho)}{1-\rho} \cdot \frac{1}{u^2} + \frac{(2 \rho^2 - 6 \rho +
7)(1+ \rho)^2}{(1-\rho)^2} \cdot \frac{1}{u^4} \right. \nonumber \\
\left.  - 3 \frac{(12-13 \rho + 8
\rho^2 - 2 \rho^3)(1+ \rho)^3}{(1-\rho)^3} \cdot \frac{1}{u^6} +  {\cal O}
\left( \frac{1}{u^8} \right) \right], \eea
and
\bea
1/L(u,u;\rho)=  \frac{2 \pi~ u^2~\sqrt{1-\rho^2}}{(1+\rho)^2}
\cdot e^{\frac{u^2}{1+\rho}} \left[ 1+ \frac{(2-\rho)(1+
\rho)}{1-\rho} \cdot \frac{1}{u^2} - \frac{3 - 2 \rho +
\rho^2)(1+\rho)^2}{(1-\rho)^2} \cdot \frac{1}{u^4} \right. \nonumber \\
\left. + \frac{(16 -13 \rho + 10 \rho^2 - 3 \rho^3)(1+\rho)^3}{(1-\rho)^3}
\cdot \frac{1}{u^6} +  {\cal O}
\left( \frac{1}{u^8} \right) \right].
\eea

\subsubsection{Asymptotic behavior of the first moment $m_{10}$}

The first moment $m_{10} = \E[X~|~X>u, Y>u]$ is given by (\ref{mfdd}).
For large $u$,
\bea
1- \Phi \left(\sqrt{\frac{1-\rho}{1+\rho}} u \right) &=&  \frac{1}{2}~
\mbox{erfc} \left(\sqrt{\frac{1-\rho}{2(1+\rho)}} u \right) \\
&=&\sqrt{ \frac{1+\rho}{1-\rho}}
\frac{e^{-\frac{1-\rho}{2(1+\rho)}u^2}}{\sqrt{2 \pi}~u} \left[1 -
\frac{1+\rho}{1-\rho} \cdot \frac{1}{u^2} + 3 \left(\frac{1+\rho}{1-\rho}
\right)^2 \cdot \frac{1}{u^4} \right. \nonumber\\
& &\left. - 15 \left(\frac{1+\rho}{1-\rho}
\right)^3 \cdot \frac{1}{u^6} + {\cal O} \left(\frac{1}{u^8} \right)
\right],
\eea
so that multiplying by $(1+\rho)~\phi(u)$, we obtain
\be
m_{10}~L(u,u;\rho) =
\frac{(1+\rho)^2}{\sqrt{1-\rho^2}}~\frac{e^{-\frac{u^2}{1+\rho}}}{2 \pi ~ u}
\left[1 - \frac{1+\rho}{1-\rho} \cdot \frac{1}{u^2} + 3
\left(\frac{1+\rho}{1-\rho} \right)^2 \cdot \frac{1}{u^4} - 15
\left(\frac{1+\rho}{1-\rho} \right)^3 \cdot \frac{1}{u^6} + {\cal O}
\left(\frac{1}{u^8} \right) \right]. \ee

Using the result given by equation (\ref{eq:Lu_as}), we can conclude that
\be
m_{10} =u +(1+\rho) \cdot \frac{1}{u} - \frac{(1+\rho)^2 (2 -\rho)}{(1- \rho)}
\cdot \frac{1}{u^3} +\frac{(10-8 \rho+3 \rho^2)(1+\rho)^3}{(1-\rho)^2}
\cdot \frac{1}{u^5} + {\cal O} \left(\frac{1}{u^7} \right). \ee

In the sequel, we will also need the behavior of ${m_{10}}^2$:
\be
{m_{10}}^2= u^2 +2~(1+\rho)  - \frac{(1+\rho)^2 (3 -\rho)}{(1- \rho)} \cdot
\frac{1}{u^2} + 2 \frac{(8-5 \rho+2 \rho^2)(1+\rho)^3}{(1-\rho)^2} \cdot
\frac{1}{u^4}+ {\cal O} \left(\frac{1}{u^6} \right). \ee

\subsubsection{Asymptotic behavior of the second moment $m_{20}$}
The second moment $m_{20}=\E[X^2~|~X>u, Y>u]$ is given by expression
(\ref{njhjjnnsw}).
The first term in the right hand side of (\ref{njhjjnnsw}) yields
\bea
(1+\rho^2)~ u~ \phi(u) \left[1-
\Phi\left(\sqrt{\frac{1-\rho}{1+\rho}} u \right) \right] = (1+\rho^2)
\sqrt{\frac{1+\rho}{1-\rho}}~\frac{e^{-\frac{u^2}{1+\rho}}}{2 \pi}
\left[1 - \frac{1+\rho}{1-\rho} \cdot \frac{1}{u^2} + 3
\left(\frac{1+\rho}{1-\rho} \right)^2 \cdot \frac{1}{u^4} \right. \nonumber\\
\left. - 15 \left(\frac{1+\rho}{1-\rho}
\right)^3 \cdot \frac{1}{u^6} + {\cal O}
\left(\frac{1}{u^8} \right) \right]~,
\eea
while the second term gives
\be
\frac{\rho~\sqrt{1-\rho^2}}{\sqrt{2 \pi}}~\phi \left( \sqrt{
\frac{2}{1+\rho}} u \right) =
\rho~\sqrt{1-\rho^2}~\frac{e^{-\frac{u^2}{1+\rho}}}{2 \pi}~.
\ee

Putting these two expressions together and factorizing the term
$(1+\rho)/(1+\rho^2)$ allows us to obtain
\bea
m_{20}~L(u,u;\rho) =\frac{(1+\rho)^2}{\sqrt{1-\rho^2}}~
\frac{e^{-\frac{u^2}{1+\rho}}}{2 \pi} \left[ 1 -
\frac{1+\rho^2}{1-\rho} \cdot \frac{1}{u^2} + 3
\frac{(1+\rho^2)(1+\rho)}{(1-\rho)^2}\cdot \frac{1}{u^4} \right. \nonumber \\
\left. - 15 \frac{(1+\rho^2)(1+\rho)^2}{(1-\rho)^3} \cdot \frac{1}{u^6} +
{\cal O} \left(\frac{1}{u^8} \right) \right] + L(u,u; \rho)~,
\eea
which finally yields
\be
m_{20} = u^2 + 2~(1+\rho) -2
\frac{(1+\rho)^2}{1- \rho} \cdot \frac{1}{u^2}+2
\frac{(5+4 \rho + \rho^3)(1+\rho)^2}{(1-\rho)^2} \frac{1}{u^4} +
{\cal O} \left(\frac{1}{u^6} \right)~.
\ee

\subsubsection{Asymptotic behavior of the cross moment $m_{11}$}
The cross moment $m_{11} = \E[X \cdot Y~|~ X>u, Y>u]$ is given by
expression (\ref{mmjhmjmr}).
The first and second terms in the right hand side of (\ref{mmjhmjmr})
respectively give
\bea
2\rho~u~\phi(u) [1- \Phi(u)]=2 \rho~
\sqrt{\frac{1+\rho}{1-\rho}}~\frac{e^{-\frac{u^2}{1+\rho}}}{2 \pi}
\left[1 - \frac{1+\rho}{1-\rho} \cdot \frac{1}{u^2} + 3
\left(\frac{1+\rho}{1-\rho} \right)^2 \cdot \frac{1}{u^4} \right. \nonumber\\
\left. - 15 \left(\frac{1+\rho}{1-\rho}
\right)^3 \cdot \frac{1}{u^6} + {\cal O}
\left(\frac{1}{u^8} \right) \right],
\eea
\be
\frac{\sqrt{1-\rho^2}}{\sqrt{2 \pi}}~\phi \left( \sqrt{
\frac{2}{1+\rho}} u \right) =
\sqrt{1-\rho^2}~\frac{e^{-\frac{u^2}{1+\rho}}}{2 \pi},
\ee
which, after factorization by $(1+\rho)/\rho$, yields
\bea
m_{11}~L(u,u;\rho) =\frac{(1+\rho)^2}{\sqrt{1-\rho^2}}~
\frac{e^{-\frac{u^2}{1+\rho}}}{2 \pi} \left[1 - 2~
\frac{\rho}{1-\rho} \cdot \frac{1}{u^2} + 6~
\frac{\rho(1+\rho)}{(1-\rho)^2}\cdot \frac{1}{u^4} \right. \nonumber\\
\left. - 30 ~ \frac{\rho (1+\rho)^2}{(1-\rho)^3} \cdot \frac{1}{u^6}+ {\cal
O} \left(\frac{1}{u^8} \right) \right] +\rho~ L(u,u; \rho),
\eea
and finally
\be
m_{11}= u^2 +2~(1+\rho)  - \frac{(1+\rho)^2 (3 -\rho)}{(1- \rho)} \cdot
\frac{1}{u^2} + \frac{(16 - 9 \rho + 3 \rho^2)(1+
\rho)^3}{(1-\rho)^2} \cdot \frac{1}{u^4} + {\cal O} \left(\frac{1}{u^6}
\right). \ee

\subsubsection{Asymptotic behavior of the correlation coefficient}

The conditional correlation coefficient conditioned on both $X$ and $Y$
larger than $u$ is defined by (\ref{ngkkekems}).
Using the symmetry between $X$ and $Y$, we have $m_{10}=m_{01}$ and
$m_{20}=m_{02}$, which allows us to rewrite (\ref{ngkkekems}) as
follows
\be
\rho_u = \frac{m_{11}-{m_{10}}^2}{m_{20}-{m_{10}}^2}~.
\ee
Putting together the previous results, we have
\bea
m_{20}-{m_{10}}^2 &=&\frac{(1+\rho)^2}{u^2} -2~ \frac{(4 - \rho + 3 \rho^2 + 3
\rho^3)(1+\rho)^2}{1-\rho} \cdot \frac{1}{u^4} + {\cal O} \left(\frac{1}{u^6}
\right),\\
m_{11}-{m_{10}}^2&=& \rho~ \frac{(1+\rho)^3}{1-\rho} \cdot
\frac{1}{u^4} +  {\cal O} \left(\frac{1}{u^6} \right), \eea
which proves that
\be
\rho_u = \rho ~ \frac{1+\rho}{1-\rho} \cdot \frac{1}{u^2}+ {\cal O}
\left(\frac{1}{u^4} \right) ~ ~ ~ ~ \mbox{and}~ ~ \rho \in [-1,1).
\ee

\newpage
%*******************************
%*              Appendix 2            *
%*******************************

\section{Conditional correlation coefficient for Student's variables}
\subsection{Proposition}
\label{app:rho_std}
Let us consider a pair of Student's random variables $(X, Y)$ with $\nu$
degrees of freedom and unconditional correlation coefficient $\rho$.
Let $\A$ be a subset of $\mathbb R$ such that $\Pr\{ Y \inA \} >0$. The
correlation coefficient of $(X,Y)$, conditioned on $Y \inA$ defined by
\be
\rho_\A  =  {\cov(X,Y~|~ Y \inA) \over \sqrt{\var(X~|~ Y \inA)}~
\sqrt{\var(Y~|~ Y \inA)}}~.
\ee
can be expressed as
\be
\label{eq:rho_std}
\rho_\A  = \frac{\rho}{\sqrt{\rho^2 + \frac{\E[
\E(x^2~|~Y)- \rho^2 Y^2~|~ Y\inA]}{\var(Y~|~Y\inA)}}}~,
\ee
with
{\small
\be
\var(Y~|~Y\inA)=\nu \left[ \frac{\nu-1}{\nu-2} \cdot
\frac{\Pr\left\{\sqrt{\frac{\nu}{\nu-2}}Y\inA~|~ \nu-2
\right\}}{\Pr\{Y\inA~|~\nu\}} -1 \right] - \left[ \frac{\int_{y\inA} dy~ y
\cdot t_y(y)}{\Pr\{Y\inA~|~ \nu\}} \right]^2,
\ee
}
and
\be
\E[\E(X^2~|~Y) - \rho^2 Y^2~|~ Y \inA] =
(1- \rho^2) ~ \frac{\nu}{\nu-2} \cdot
\frac{\Pr\left\{\sqrt{\frac{\nu}{\nu-2}}Y\inA~|~ \nu-2
\right\}}{\Pr\{Y\inA~|~\nu\}}~.
\ee

\subsection{Proof of the proposition}
Let the variables $X$ and $Y$ have a multivariate Student's distribution with
$\nu$ degrees of freedom and a correlation coefficient $\rho$ :
\bea
P_{XY}(x,y) &=& \frac{\Gamma \left(\frac{\nu+2}{2} \right)}{\nu \pi ~ \Gamma
\left(\frac{\nu+1}{2} \right) \sqrt{1-\rho^2}}~
\left(1+\frac{ x^2-2\rho x y+y^2}{\nu~(1-\rho^2)}  \right) ^{-\frac{\nu+2}{2}}
,\\
&=& \left(\frac{\nu+1}{\nu+y^2}\right)^{1/2} \frac{1}{
\sqrt{1- \rho^2}}  ~t_\nu(y) \cdot
t_{\nu+1} \left[ \left( \frac{\nu+1}{\nu+y^2}\right)^{1/2} \frac{x-\rho y}{
\sqrt{1- \rho^2}}  \right],
\label{eq:std_fac}
\eea
where $t_\nu(\cdot)$ denotes the univariate
Student's density with $\nu$ degrees of freedom
\be
t_\nu(x)=\frac{\Gamma \left(\frac{\nu+1}{2} \right)}{\Gamma
\left(\frac{\nu}{2} \right) (\nu \pi)^{1/2}} \cdot \frac{1}{\left(1 +
\frac{x^2}{\nu} \right)^\frac{\nu+1}{2}}= \frac{C_\nu}{\left(1 +
\frac{x^2}{\nu} \right)^\frac{\nu+1}{2}}~.
\label{mgjk}
\ee

Let us evaluate $\cov(X,Y~|~ Y \inA)$:
\bea
\cov(X,Y~|~ Y \inA) &=& \E(X\cdot Y~|~ Y \inA)-\E(X~|~ Y \inA) \cdot \E(Y~|~ Y
\inA),\\
&=& \E(\E(X~|~Y)\cdot Y~|~ Y \inA)-\E(\E(X~|~Y)~|~ Y \inA) \cdot \E(Y~|~ Y
\inA).
\eea
As it can be seen in equation (\ref{eq:std_fac}),
$\E(X~|~Y)=\rho Y$, which gives \bea
\cov(X,Y~|~ Y \inA) &=&\rho \cdot \E( Y^2~|~ Y \inA)-\rho \cdot \E(Y~|~ Y
\inA)^2,\\
&=& \rho \cdot \var( Y~|~ Y \inA).
\eea
Thus, we have
\be
\rho_\A= \rho \sqrt{\frac{\var( Y~|~ Y \inA)}{\var( X~|~ Y \inA)}}.
\ee

Using the same method as for the calculation of $\cov(X,Y~|~ Y \inA)$, we find
\bea
\var(X~|~Y \inA)&=&\E[\E(X^2~|~Y)~|~Y \inA)] - \E[\E(X~|~Y)~|~Y \inA)]^2,\\
&=& \E[\E(X^2~|~Y)~|~Y \inA)] -\rho^2\cdot \E[Y~|~Y \inA]^2,\\
&=& \E[\E(X^2~|~Y) - \rho^2 Y^2~|~Y \inA)] -\rho^2\cdot \var[Y~|~Y \inA],\\
\eea
which yields
\be
\rho_\A  = \frac{\rho}{\sqrt{\rho^2 + \frac{\E[
\E(x^2~|~Y)- \rho^2 Y^2~|~ Y\inA]}{\var(Y~|~Y\inA)}}},
\ee
as asserted in (\ref{eq:rho_std}).

To go one step further, we have to evaluate the three terms $\E(Y~|~ Y \inA)$,
$\E(Y^2~|~ Y \inA)$,  and $\E[ \E(X^2~|~Y)~|~ Y \inA]$.

The first one is trivial to calculate :
\be
\E(Y~|~ Y \inA) = \frac{\int_{y\inA} dy~ y \cdot t_y(y)}{\Pr\{Y\inA~|~ \nu\}}.
\ee

The second one gives
\bea
\E(Y^2~|~ Y \inA) &=& \frac{\int_{y\inA} dy~ y^2 \cdot t_y(y)}{\Pr\{Y\inA~|~
\nu\}},\\
\label{eq:e2}
&=& \nu \left[ \frac{\nu-1}{\nu-2} \cdot
\frac{\Pr\left\{\sqrt{\frac{\nu}{\nu-2}}Y\inA~|~ \nu-2
\right\}}{\Pr\{Y\inA~|~\nu\}} -1 \right],
\eea
so that
\be
\var(Y~|~Y\inA)=\nu \left[ \frac{\nu-1}{\nu-2} \cdot
\frac{\Pr\left\{\sqrt{\frac{\nu}{\nu-2}}Y\inA~|~ \nu-2
\right\}}{\Pr\{Y\inA~|~\nu\}} -1 \right] - \left[ \frac{\int_{y\inA} dy~ y
\cdot t_y(y)}{\Pr\{Y\inA~|~ \nu\}} \right]^2.
\ee

To calculate the third term, we first need to evaluate $\E(X^2~|~Y)$. Using
equation (\ref{eq:std_fac}) and the results given in \cite{Abra}, we find
\bea
\E(X^2~|~Y) &=& \int dx~    \left(\frac{\nu+1}{\nu+y^2}\right)^{1/2}
\frac{x^2}{ \sqrt{1- \rho^2}}  \cdot
t_{\nu+1} \left[ \left( \frac{\nu+1}{\nu+y^2}\right)^{1/2} \frac{x-\rho y}{
\sqrt{1- \rho^2}}  \right],
\\
&=& \frac{\nu+y^2}{\nu-1}~(1-\rho^2) - \rho^2 y^2~,
\label{mgxckke}
\eea
which yields
\be
\E[\E(X^2~|~Y) - \rho^2 Y^2~|~ Y \inA] = \frac{\nu}{\nu-1} (1-\rho^2) +
\frac{1-\rho^2}{\nu-1} ~ \E[Y^2~|~ Y \inA]~,
\ee
and applying the result given in eqation (\ref{eq:e2}), we finally obtain
\be
\E[\E(X^2~|~Y) - \rho^2 Y^2~|~ Y \inA] =
(1- \rho^2) ~ \frac{\nu}{\nu-2} \cdot
\frac{\Pr\left\{\sqrt{\frac{\nu}{\nu-2}}Y\inA~|~ \nu-2
\right\}}{\Pr\{Y\inA~|~\nu\}},
\ee
which concludes the proof.

\subsection{Conditioning on $Y$ larger than $v$}
\label{app:os_ccc_sv}
The conditioning set is $\A = [v, +\infty)$, thus
\bea
\label{eq:P1}
\Pr\{Y\inA~|~\nu\} &=& \bar T_\nu(v) =\nu^\frac{\nu-1}{2}~ \frac{ C_\nu}{v^\nu}
+ {\cal O}\left(v^{-(\nu+2)}\right), \\
\label{eq:P2}
\Pr\left\{\sqrt{\frac{\nu}{\nu-p}}Y\inA~|~ \nu-p
\right\} &=& \bar T_{\nu-p}  \left( \sqrt{\frac{\nu-p}{\nu}}~ v \right)
=\frac{\nu ^\frac{\nu-p}{2}}{(\nu-p)
^\frac{1}{2}}~ \frac{ C_{\nu-p}}{v^{\nu-p}} + {\cal
O}\left(v^{-(\nu-p+2)}\right),\\
\int_{y\inA} dy~ y \cdot t_y(y) &=& \sqrt{ \frac{\nu}{\nu-2}}~ t_{\nu-2}
\left( \sqrt{\frac{\nu-2}{\nu}}~ v \right) =
\frac{\nu^\frac{\nu}{2}}{\sqrt{\nu-2}}~ \frac{C_{\nu-2}}{v^{\nu-1}} + {\cal
O}\left(v^{-(\nu-3)}\right), \eea
where $t_\nu(\cdot)$ and  $\bar T_\nu(\cdot)$ denote respectively the density
and the Student's survival distribution with $\nu$ degrees of freedom and
$C_\nu$ is defined in (\ref{mgjk}).

Using equation (\ref{eq:rho_std}), one can thus give the exact expression of
$\rho_v^+$. Since it is very cumbersomme, we will not write it explicitely. We
will only give the asymptotic expression of $\rho_v^+$. In this respect, we can
show that
\bea
\var(Y~|~Y\inA)&=& \frac{\nu}{(\nu-2)(\nu-1)^2} ~v^2 + {\cal O}(1)
\label{ngvcvgz}\\
\E[\E(X^2~|~Y) - \rho^2
Y^2~|~ Y \inA]&=& \sqrt {\frac{\nu}{\nu-2}} \frac{1-\rho^2}{\nu-1}~ v^2
+ {\cal O}(1)~.
\label{eq:P3}
\eea
Thus, for large $v$,
\be
\rho_v^+ \longrightarrow \frac{\rho}{\sqrt{  \rho^2 + (\nu-1)
\sqrt{\frac{\nu-2}{\nu}}~ (1-\rho^2)}}.
\ee

\subsection{Conditioning on $|Y|$ larger than $v$}
\label{app:os_ccc_sv2}
The conditioning set is now $\A=(- \infty, -v] \cup [v, +\infty)$, with $v
\in {\mathbb R_+}$. Thus, the right hand sides of equations (\ref{eq:P1})
and (\ref{eq:P2}) have to be multiplied by two while
\be
\int_{y\inA} dy~ y \cdot t_y(y)=0,
\ee
for symmetry reasons. So the equation (\ref{eq:P3}) still holds while
\be
\var(Y~|~Y\inA)= \frac{\nu}{(\nu-2)} ~v^2 + {\cal O}(1)~.
\label{nvbgxts}
\ee
Thus, for large $v$,
\be
\rho_v^s \longrightarrow \frac{\rho}{\sqrt{  \rho^2 + \frac{1}{(\nu-1)}
\sqrt{\frac{\nu-2}{\nu}}~ (1-\rho^2)}}.
\ee

%*******************************
%*              Appendix 3            *
%*******************************

\newpage

\section{Proof of equation (\ref{eq:rho_fac})}
\label{app:rho_fac}

We assume that $X$ and $Y$ are related by the equation
\be
X= \alpha Y + \epsilon~,
\ee
where $\alpha$ is a non random real coefficient and $\epsilon$ an idiosyncratic
noise independent of $Y$, whose distribution is assumed to admit a moment of
second order $\sigma_\epsilon ^2$. Let us also denote by $\sigma_y^2$ the
second moment of the variable $Y$.

We have
\bea
\cov(X,Y~|~ Y \inA) &=& \cov(\alpha Y + \epsilon, Y~|~ Y \inA),\\
&=& \alpha \var(Y~|~ Y \inA) + \cov(\epsilon, Y~|~ Y \inA),\\
&=& \alpha  \var(Y~|~ Y \inA),
\eea
since $Y$ and $\epsilon$ are independent.
We have also
\bea
\var(X~|~ Y \inA) &=& =\alpha^2 \var(Y~|~ Y\inA) + 2~\cov(\epsilon, Y~|~ Y
\inA) + \var(\epsilon ~|~ Y \inA),\\
&=& \alpha^2 \var(Y~|~ Y\inA) + \sigma_\epsilon^2,
\eea
where, again, we have used the independence of $Y$ and $\epsilon$.
This allows us to write
\bea
\rho_\A &=& \frac{\alpha  \var(Y~|~ Y \inA)}{\sqrt{\var(Y~|~ Y\inA)(\alpha^2
\var(Y~|~ Y\inA) + \sigma_\epsilon^2)}},\\
&=& \frac{\mbox{sgn}(\alpha)}{\sqrt{1+\frac{\sigma_\epsilon^2}{\alpha^2}\cdot
\frac{1}{\var(Y~|~Y\inA)}}}.
\eea
Since
\be
\rho= \frac{\mbox{sgn}(\alpha)}{\sqrt{1+\frac{\sigma_\epsilon^2}{\alpha^2}\cdot
\frac{1}{\var(Y)}}},
\ee
we finally obtain
\be
\rho_\A = \frac{\rho}{\sqrt{\rho^2 + (1-\rho^2)
\frac{\var(y)}{\var(y~|~y\inA)}}},
\ee
which conclude the proof.

\newpage

%****************************
%*                Appendix 4             *
%****************************

\section{Conditional Spearman's rho}
\label{app:sp}

The conditional Spearman's rho has been defined by
\be
\rho_s(\v) =  \frac{\cov(U,V~|~V \ge \v)}{\sqrt{\var(U~|~ V \ge \v)
\var(V~|~ \ge \v)}},
\ee

We have
\be
\E[ \cdot~|~V \ge \v] = \frac { \int_\v^1 \int_0 ^1 ~ \cdot~ dC(u,v)  }{
\int_\v^1 \int_0^1 dC(u,v)} = \frac{1}{1-\v} \int_\v^1 \int_0 ^1 ~ \cdot~
dC(u,v) ~, \ee
thus, performing a simple integration by parts, we obtain
\bea
\E[U~|~V \ge \v] &=& 1+ \frac{1}{1-\v} \left[ \int_0^1 du ~ C(u, \v) -
\frac{1}{ 2}\right]~,\\
\E[V~|~V \ge \v] &=& \frac{1+\v}{2}~,\\
\E[U^2~|~V \ge \v] &=& 1+ \frac{2}{1-\v} \left[ \int_0^1 du ~u~ C(u, \v) -
\frac{1}{ 3}\right]~,\\
\E[V^2~|~V \ge \v] &=& \frac{\v^2+\v+1}{3}~,\\
\E[U \cdot V~|~V \ge \v] &=& \frac{1+\v}{2} + \frac{1}{1-\v} \left[ \int_\v^1
dv \int_0^1 du~C(u,v) + \v~\int_0^1 du ~ C(u,\v)  - \frac{1}{2} \right]~,
\eea
which yields
\bea
\cov(U,V~|~V \ge \v) &=& \frac{1}{1-\v} \int_\v^1 dv \int_0^1 du~C(u,v) -
\frac{1}{2}  \int_0^1 du
~ C(u,\v) - \frac{1}{4}~, \\
\var(U~|~ V \ge \v) &=&  \frac{1-4 \v}{12~ (1-\v)^2} + \frac{2}{1-\v} \int_0^1
du~ u~C(u,\v) + \frac{2 \v-1}{(1-\v)^2} \int_0^1 du ~ C(u,\v) \nonumber \\
  & &- \frac{1}{(1-\v)^2} \left( \int_0^1 du ~ C(u,\v) \right)^2 ~, \\
  \var(V~|~ V \ge
\v) &=& \frac{(1-\v)^2}{12}~,
\eea
so that
\be
\rho_s(\v) =\frac{\frac{12}{1-\v} \int_\v^1 dv \int_0^1 du~C(u,v) -
6 \int_0^1 du
~ C(u,\v) - 3}{\sqrt{
1-4 \v + 24~(1-\v) \int_0^1
du~ u~C(u,\v) + 12~( 2\v-1) \int_0^1 du ~ C(u,\v)
- 12 \left( \int_0^1 du ~ C(u,\v) \right)^2
}}
\ee

%*******************************
%*              Appendix 5            *
%*******************************

\section{Tail dependence generated by the Student's factor model}
\label{app:tls}

We consider two random variables $X$ and $Y$, related by the relation
\be
X=\alpha Y+ \epsilon,
\ee
where $\epsilon$ is a random variable independent of $Y$ and $\alpha$ a non
random positive coefficient. Assume that $Y$ and $\epsilon$ have a Student's
distribution with density:
\bea
P_Y(y)=\frac{C_\nu}{\left( 1+ \frac{y^2}{\nu} \right) ^\frac{\nu+1}{2}}~,
\label{mgjkjer}\\
P_\epsilon(\epsilon)=\frac{C_\nu}{\sigma \left( 1+
\frac{\epsilon^2}{\nu~\sigma^2} \right) ^\frac{\nu+1}{2}}. \label{eq:pe}
\eea

We first give a general expression for the probability for $X$ to be
larger than
$F_X^{-1}(u)$ knowing that $Y$ is larger than $ F_Y^{-1}(u)$~:
\begin{lemma}
\label{lem:1}
The probability that $X$ is larger than $F_X^{-1}(u)$ knowing that $Y$ is
larger than $ F_Y^{-1}(u)$ is given by~:
\be
\label{eq:1}
\Pr[ X > F_X^{-1}(u) | Y > F_Y^{-1}(u)] =
\bar F_\epsilon(\eta) + \frac{\alpha}{1-u} \int_{F_Y^{-1}(u)}^{\infty}
dy ~\bar F_Y(y) \cdot
P_\epsilon [\alpha F_Y^{-1}(u)+\eta-\alpha y]~,
\ee
with
\be
\label{eq:eta}
\eta=F_X^{-1}(u)-\alpha F_Y^{-1}(u).
\ee
\end{lemma}
The proof of this lemma relies on a simple integration by part and a change of
variable, which are detailed in appendix \ref{app:lem1}.

Introducing the notation
\be
\tilde Y_u = F_Y^{-1}(u)~,   \label{mgmjkeees}
\ee
we can show that
\be
\label{eq:eta_y}
\eta = \alpha \left[ \left( 1+ \left( \frac{\sigma}{\alpha} \right) ^\nu
\right) ^{1/\nu}-1  \right] \tilde Y_u + {\cal O}( \tilde Y_u ^{-1} ),
\ee
which allows us to conclude that $\eta$ goes to infinity as $u$ goes to 1 (see
appendix \ref{app:eq_eta} for the derivation of this result). Thus, $\bar
F_\epsilon(\eta)$ goes to zero as u goes to 1 and
\be
\label{eq:lim_l}
\lambda = \lim_{u \rightarrow 1}  \frac{\alpha}{1-u}
\int_{\tilde Y_u}^{\infty} dy ~\bar F_Y(y) \cdot
P_\epsilon(\alpha  \tilde Y_u+\eta-\alpha y)~.
\ee

Now, using the following result~:
\begin{lemma}
\label{lem:2}
Assuming $\nu>0$ and  $x_0>1$,
\be
\lim_{\epsilon \rightarrow 0} \frac{1}{\epsilon}\int_1^\infty
dx~\frac{1}{x^\nu} \frac{C_\nu}{\left[1+\left(\frac{x-x_0}{\epsilon} \right)^2
\right]^\frac{\nu+1}{2}} = \frac{1}{x_0^\nu}, \ee
\end{lemma}
whose proof is given in appendix \ref{app:lem2}, it is straigthforward to show
that
\be
\label{eq:lamb_stud}
\lambda = \frac{1}{1+ \left( \frac{\sigma}{\alpha} \right) ^\nu}.
\ee
The final steps of this calculation are given in appendix \ref{app:eq_ls}.

\subsection{Proof of Lemma \ref{lem:1}}
\label{app:lem1}

By definition,
\bea
\Pr[ X > F_X^{-1}(u) , Y > F_Y^{-1}(u)] &=& \int_{F_X^{-1}(u)}^\infty dx
\int_{F_Y^{-1}(u)}^\infty dy  ~ P_Y(y)\cdot P_\epsilon(x-\alpha y)\\
&=&   \int_{F_Y^{-1}(u)}^\infty dy  ~ P_Y(y) \cdot \bar
F_\epsilon[F_X^{-1}(u)-\alpha y] .
\eea
Let us perform an integration by part~:
\bea
\Pr[ X > F_X^{-1}(u) , Y > F_Y^{-1}(u)] &=& \left[ - \bar F_Y(y) \cdot \bar
F_\epsilon ( F_X^{-1}(u)-\alpha y) \right]_{F_Y^{-1}(u)}^\infty
+\nonumber\\
&+& \alpha \int_{F_Y^{-1}(u)}^\infty dy~\bar F_Y(y) \cdot
P_\epsilon(F_X^{-1}(u)-\alpha y)\\
&=& (1-u) \bar F_\epsilon(F_X^{-1}(u)-\alpha F_Y^{-1}(u))
+\nonumber\\
&+& \alpha \int_{F_Y^{-1}(u)}^\infty dy~\bar F_Y(y) \cdot
P_\epsilon(F_X^{-1}(u)-\alpha y)
\eea
Defining $\eta=F_X^{-1}(u)-\alpha F_Y^{-1}(u)$ (see equation(\ref{eq:eta})),
and dividing each term by \be
\Pr[Y > F_Y^{-1}(u)]=1-u,
\ee
we obtain the result given in  (\ref{eq:1})

\subsection{Derivation of equation (\ref{eq:eta_y})}
\label{app:eq_eta}
The factor $Y$ and the idiosyncratic noise $\epsilon$ have Student's
distributions with $\nu$ degrees of freedom given by (\ref{mgjkjer})
and (\ref{eq:pe}) respectively.
It follows that the survival distributions of $Y$ and $\epsilon$ are~:
\bea
\bar F_Y(y)=\frac{\nu^\frac{\nu-1}{2}~C_\nu}{ y^\nu}+ {\cal
O}(y^{-(\nu+2)}), \label{eq:fy}\\ \bar
F_\epsilon(\epsilon)=\frac{\sigma^\nu~\nu^\frac{\nu-1}{2}~C_\nu}{\epsilon^\nu}+
{\cal O}(\epsilon^{-(\nu+2)}),\\ \eea and
\be
\label{eq:bfx}
\bar F_X(x)=\frac{(\alpha^\nu+\sigma^\nu)~\nu^\frac{\nu-1}{2}~C_\nu}{ 
x^\nu}+ {\cal
O}(x^{-(\nu+2)}). \ee

Using the notation (\ref{mgmjkeees}), equation (\ref{eq:eta})
can be rewritten as
\be
\bar F_X(\eta + \alpha \tilde Y_u) = \bar F_Y( \tilde Y_u) =1- u,
\ee
whose solution for large $\tilde Y_u$ (or equivalently as u goes to 1) is
\be
\eta = \alpha \left[ \left( 1+ \left( \frac{\sigma}{\alpha} \right) ^\nu
\right) ^{1/\nu}-1  \right] \tilde Y_u + {\cal O}( \tilde Y_u ^{-1} ).
\ee
To obain this equation, we have used the asymptotic expressions of $\bar F_X$
and $\bar F_Y$ given in (\ref{eq:bfx}) and (\ref{eq:fy}).

\subsection{ Proof of lemma \ref{lem:2}}
\label{app:lem2}
We want to prove that, assuming  $\nu>0$ and  $x_0>1$,
\be
\lim_{\epsilon \rightarrow 0} \frac{1}{\epsilon}\int_1^\infty
dx~\frac{1}{x^\nu}
\frac{C_\nu}{\left[1+\frac{1}{\nu}~\left(\frac{x-x_0}{\epsilon} \right)^2
\right]^\frac{\nu+1}{2}} = \frac{1}{x_0^\nu}. \ee

The change of variable
\be
u=\frac{x-x_0}{\epsilon}~,
\ee
gives
\bea
\frac{1}{\epsilon}\int_1^\infty
dx~\frac{1}{x^\nu} 
\frac{C_\nu}{\left[1+\frac{1}{\nu}~\left(\frac{x-x_0}{\epsilon} 
\right)^2
\right]^\frac{\nu+1}{2}} &=& \int_\frac{1-x_0}{\epsilon}^\infty du
\frac{1}{(\epsilon u +x_0)^\nu}
\frac{C_\nu}{(1+\frac{u^2}{\nu})^\frac{\nu+1}{2}} \\ &=& \frac{1}{x_0^\nu}
\int_\frac{1-x_0}{\epsilon}^\infty du \frac{1}{(1+ \frac{\epsilon u}{x_0})^\nu}
\frac{C_\nu}{(1+\frac{u^2}{\nu})^\frac{\nu+1}{2}}\\
&=&  \frac{1}{x_0^\nu}
\int_\frac{1-x_0}{\epsilon}^\frac{x_0}{\epsilon} du \frac{1}{(1+ \frac{\epsilon
u}{x_0})^\nu} \frac{C_\nu}{(1+\frac{u^2}{\nu})^\frac{\nu+1}{2}} + \nonumber \\
&+& \frac{1}{x_0^\nu} \int_\frac{x_0}{\epsilon}^\infty du \frac{1}{(1+
\frac{\epsilon u}{x_0})^\nu}
\frac{C_\nu}{(1+\frac{u^2}{\nu})^\frac{\nu+1}{2}}~. \eea

Consider the second integral. We have
\be
u \ge \frac{x_0}{\epsilon},
\ee
which allows us to write
\be
   \frac{1}{(1+u^2)^\frac{\nu+1}{2}} \le
\frac{\nu^\frac{\nu+1}{
2}~\epsilon^{\nu+1}}{x_0^{\nu+1}}, \ee
so that
\bea
\left| \int_\frac{x_0}{\epsilon}^\infty du
\frac{1}{(1+ \frac{\epsilon u}{x_0})^\nu} \frac{C_\nu}{(1+u^2)^\frac{\nu+1}{2}}
\right|  &\le&  \frac{\nu^\frac{\nu+1}{2}~\epsilon^{\nu+1}}{x_0^{\nu+1}}
\int_\frac{x_0}{\epsilon}^\infty du \frac{C_\nu}{(1+ \frac{\epsilon
u}{x_0})^\nu}\\ &=&  \frac{\nu^\frac{\nu+1}{2}~\epsilon^{\nu}}{x_0^{\nu}}
\int_1^\infty dv \frac{C_\nu}{(1+ v)^\nu}\\
&=& {\cal O}(\epsilon^\nu).   \label{eq:20}
\eea

The next step of the proof is to show that
\be
\int_\frac{1-x_0}{\epsilon}^\frac{x_0}{\epsilon} du \frac{1}{(1+ \frac{\epsilon
u}{x_0})^\nu} \frac{C_\nu}{(1+\frac{u^2}{\nu})^\frac{\nu+1}{2}}
\longrightarrow 1~ ~ ~ \mbox{as} ~ ~ ~ \epsilon \longrightarrow 0. \ee

Let us calculate
\bea
\left| \int_\frac{1-x_0}{\epsilon}^\frac{x_0}{\epsilon} du
\frac{1}{(1+ \frac{\epsilon
u}{x_0})^\nu} \frac{C_\nu}{(1+\frac{u^2}{\nu})^\frac{\nu+1}{2}} - 1 \right|
&=& \left| \int_\frac{1-x_0}{\epsilon}^\frac{x_0}{\epsilon} du \frac{1}{(1+
\frac{\epsilon u}{x_0})^\nu} \frac{C_\nu}{(1+\frac{u^2}{\nu})^\frac{\nu+1}{2}}
- \right. \nonumber \\
&-& \left. \int_{-\infty}^\infty du
\frac{C_\nu}{(1+\frac{u^2}{\nu})^\frac{\nu+1}{2}} \right|\\
&=&    \left| \int_\frac{1-x_0}{\epsilon}^\frac{x_0}{\epsilon} du \left[
\frac{1}{(1+ \frac{\epsilon u}{x_0})^\nu} -1 \right]
\frac{C_\nu}{(1+\frac{u^2}{\nu})^\frac{\nu+1}{2}} - \right. \nonumber\\
&-&\left. \int_{-\infty}^\frac{1-x_0}{\epsilon} du
\frac{C_\nu}{(1+\frac{u^2}{\nu})^\frac{\nu+1}{2}} -
\int_\frac{x_0}{\epsilon}^\infty du
\frac{C_\nu}{(1+\frac{u^2}{\nu})^\frac{\nu+1}{2}} \right|\\ &\le& \left|
\int_\frac{1-x_0}{\epsilon}^\frac{x_0}{\epsilon} du \left[ \frac{1}{(1+
\frac{\epsilon u}{x_0})^\nu} -1 \right]
\frac{C_\nu}{(1+\frac{u^2}{\nu})^\frac{\nu+1}{2}} \right| + \nonumber\\
&+& \left| \int_{-\infty}^\frac{1-x_0}{\epsilon} du
\frac{C_\nu}{(1+\frac{u^2}{\nu})^\frac{\nu+1}{2}} \right| + \left|
\int_\frac{x_0}{\epsilon}^\infty du
\frac{C_\nu}{(1+\frac{u^2}{\nu})^\frac{\nu+1}{2}} \right|. \eea

The second and third integrals obviously behave like ${\cal O}(\epsilon^\nu)$
when $\epsilon$ goes to zero since we have assumed $x_0>1$ what ensures that
$\frac{1-x_0}{\epsilon} \rightarrow - \infty$ and $\frac{x_0}{\epsilon}
\rightarrow \infty$ when $\epsilon \rightarrow 0$. For the first integral, we
have
\be \left| \int_\frac{1-x_0}{\epsilon}^\frac{x_0}{\epsilon} du \left[
\frac{1}{(1+ \frac{\epsilon u}{x_0})^\nu} -1 \right]
\frac{C_\nu}{(1+\frac{u^2}{\nu})^\frac{\nu+1}{2}} \right| \le
\int_\frac{1-x_0}{\epsilon}^\frac{x_0}{\epsilon} du \left| \frac{1}{(1+
\frac{\epsilon u}{x_0})^\nu} -1 \right|
\frac{C_\nu}{(1+\frac{u^2}{\nu})^\frac{\nu+1}{2}} . \ee

\begin{figure}
\begin{center}
\includegraphics[width=15cm]{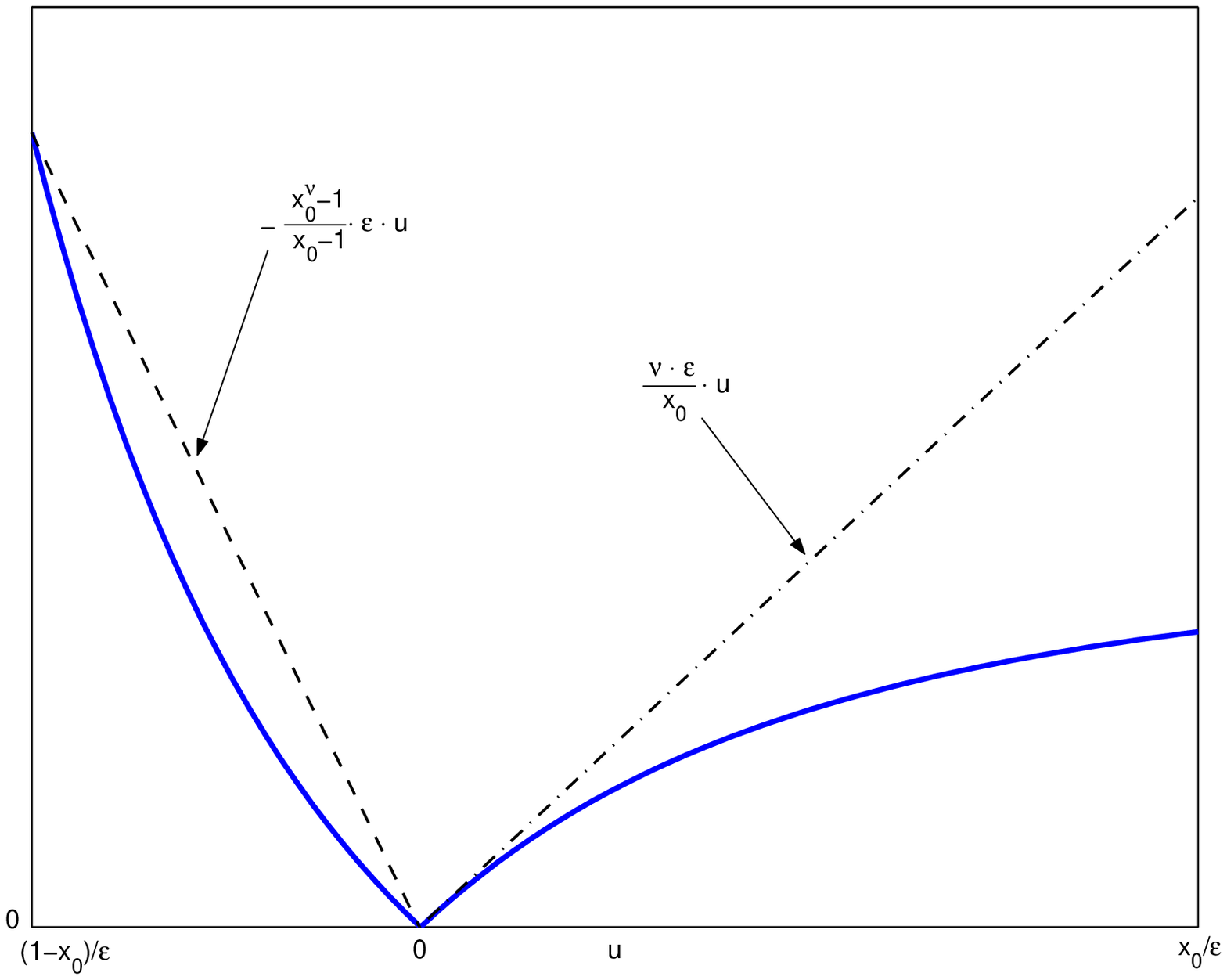}
\caption{\label{fig:majoration} The graph of the function
$ \left| \frac{1}{(1+ \frac{\epsilon u}{x_0})^\nu} -1 \right| $ (thick solid
line), the string which gives an upper bound of the function within $\left[
\frac{1-x_0}{\epsilon},0 \right]$ (dashed line) and the tangent in $0^+$
which gives an upper bound of the function within $\left[0,
\frac{x_0}{\epsilon} \right]$ (dash dotted line).}
\end{center}
\end{figure}

The function
\be
\left| \frac{1}{(1+
\frac{\epsilon u}{x_0})^\nu} -1 \right|
\ee
vanishes at $u=0$, is convex for $u \in [ \frac{1-x_0}{\epsilon},0]$ and
concave for $u \in [0, \frac{x_0}{\epsilon}]$
(see also figure
\ref{fig:majoration}), so that  there are two constants $A,B>0$ such that
\bea
\left| \frac{1}{(1+
\frac{\epsilon u}{x_0})^\nu} -1 \right| &\le& -\frac{x_0^\nu-1}{x_0-1} \epsilon
\cdot u =- A \cdot \epsilon \cdot u , ~~~~\forall  u \in \left[
\frac{1-x_0}{\epsilon},0 \right] \\ \left| \frac{1}{(1+
\frac{\epsilon u}{x_0})^\nu} -1 \right| &\le& \frac{\nu \epsilon}{x_0}
   u = B \cdot \epsilon \cdot u , ~~~~\forall  u \in \left[0,
\frac{x_0}{\epsilon} \right] . \eea
We can thus conclude that
\bea
\left|
\int_\frac{1-x_0}{\epsilon}^\frac{x_0}{\epsilon} du \left[ \frac{1}{(1+
\frac{\epsilon u}{x_0})^\nu} -1 \right]
\frac{C_\nu}{(1+\frac{u^2}{\nu})^\frac{\nu+1}{2}} \right| &\le& -A \cdot
\epsilon \int_\frac{1-x_0}{\epsilon}^0 du \frac{u \cdot
C_\nu}{(1+\frac{u^2}{\nu})^\frac{\nu+1}{2}} \nonumber\\ &+& B \cdot \epsilon
\int_0^\frac{x_0}{\epsilon} du \frac{u \cdot
C_\nu}{(1+\frac{u^2}{\nu})^\frac{\nu+1}{2}}\\ &=& {\cal O}(\epsilon^\alpha),
\eea
with $\alpha = \min \{ \nu, 1\}$. Indeed, the two integrals can be perfomed
exactly, which shows that they behave as ${\cal O}(1)$ if $\nu >1$ and as
${\cal O}( \epsilon^{\nu-1})$ otherwise. Thus, we finally obtain
\be
\label{eq:30}
\left| \int_\frac{1-x_0}{\epsilon}^\frac{x_0}{\epsilon} du
\frac{1}{(1+ \frac{\epsilon
u}{x_0})^\nu} \frac{C_\nu}{(1+\frac{u^2}{\nu})^\frac{\nu+1}{2}} - 1 \right| =
{\cal O} (\epsilon ^\alpha).
\ee
Putting together equations (\ref{eq:20}) and (\ref{eq:30}) we obtain
\be
\left|\frac{1}{\epsilon}\int_1^\infty
dx~\frac{1}{x^\nu}
\frac{C_\nu}{\left[1+\frac{1}{\nu}~\left(\frac{x-x_0}{\epsilon}
\right)^2 \right]^\frac{\nu+1}{2}} - \frac{1}{x_0^\nu} \right| ={\cal
O}(\epsilon^{\min \{ \nu, 1\}})~, \ee
which concludes the proof.

\subsection{Derivation of equation (\ref{eq:lamb_stud})}
\label{app:eq_ls}
  From equation (\ref{eq:fy}), we can deduce
\be
\bar F_Y(y)=\frac{\nu^\frac{\nu-1}{2}~C_\nu}{ y^\nu}\left( 1+ {\cal 
O}(y^{-2}) \right).
\ee

Using equations (\ref{eq:pe}) and (\ref{eq:eta_y}), we obtain
\be
P_\epsilon(\alpha  \tilde Y_u+\eta-\alpha y) = P_\epsilon (\gamma \tilde Y_u
-\alpha y) \cdot \left( 1+ {\cal O}(\tilde Y_u^{-2}) \right),
\ee
where
\be
\gamma =    \alpha  \left( 1+ \left( \frac{\sigma}{\alpha} \right) ^\nu
\right) ^{1/\nu} .
\ee

Putting together these results yields for the leading order
\bea
\int_{\tilde Y_u}^{\infty} dy ~\bar F_Y(y) \cdot
P_\epsilon(\alpha  \tilde Y_u+\eta-\alpha y) &=& \int_{\tilde Y_u}^{\infty} dy
~ \frac{\nu^\frac{\nu-1}{2}~C_\nu}{ y^\nu}\cdot  \frac{C_\nu}{\sigma \left( 1+
\frac{(\gamma \tilde Y_u
-\alpha y)^2}{\nu~\sigma^2} \right) ^\frac{\nu+1}{2}},\\
&=&\frac{\nu^\frac{\nu-1}{2}~C_\nu}{\alpha~\tilde {Y_u}^\nu} \int_1^{\infty} dx
~ \frac{1}{x^\nu}\cdot  \frac{C_\nu ~ \frac{\alpha \tilde Y_u}{\sigma}}{
\left( 1+ \frac{1}{\nu} \left(
\frac{x-\frac{\gamma}{\alpha}}{\frac{\sigma}{\alpha \tilde Y_u}} \right)^2
\right) ^\frac{\nu+1}{2}}, \eea where the change of variable $x=\frac{y}{\tilde
Y_u}$ has been performed in the last equation.

We now apply lemma \ref{lem:2} with $x_0= \frac{\gamma}{\alpha}>1$ and
$\epsilon =\frac{\sigma}{\alpha \tilde Y_u}$ which goes to zero as u
goes to 1. This gives
\be
\int_{\tilde Y_u}^{\infty} dy ~\bar F_Y(y) \cdot
P_\epsilon(\alpha  \tilde Y_u+\eta-\alpha y) \sim_{u \rightarrow 1}
\frac{\nu^\frac{\nu-1}{2}~C_\nu}{\alpha~{\tilde Y_u}^\nu}~  \left(
\frac{\alpha}{\gamma} \right)^\nu~,
\ee
which shows that
\be
\Pr[ X > F_X^{-1}(u) , Y > F_Y^{-1}(u)] \sim_{u \rightarrow 1}
F_Y^{-1}( \tilde Y_u)~  \left( \frac{\alpha}{\gamma}
\right)^\nu = (1-u)~  \left( \frac{\alpha}{\gamma}
\right)^\nu,
\ee
thus
\be
\Pr[ X > F_X^{-1}(u) | Y > F_Y^{-1}(u)] \sim_{u \rightarrow 1}
  \left( \frac{\alpha}{\gamma}
\right)^\nu, \ee
which finally yields
\be
\lambda = \frac{1}{1+ \left( \frac{\sigma}{\alpha} \right) ^\nu}.
\ee

\newpage

\begin{table}
\begin{center}
\begin{tabular}{||c|c|c|c||}
\hline
\hline
& $\rho_{v}^+$ & $\rho_{v}^s$ &$\rho_{u}$ \\
\hline
Bivariate Gaussian & $\frac{\rho}{\sqrt{1-\rho^2}}\cdot
\frac{1}{v}$~~(\ref{eq:os_ccc_gv}) & $1-
{1 \over 2}~{1-\rho^2 \over \rho^2}~{1 \over
v^2}$~~(\ref{eq:os_ccc_gv2}) & $\rho \frac{1+\rho}{1-\rho} \cdot
\frac{1}{u^2}$~~(\ref{mgjzma})  \\ \hline
Bivariate Student's  & $\frac{\rho}{\sqrt{  \rho^2 + (\nu-1)
\sqrt{\frac{\nu-2}{\nu}}~ (1-\rho^2)}}$~~(\ref{mhytl})
& $\frac{\rho}{\sqrt{  \rho^2 + \frac{1}{(\nu-1)}
\sqrt{\frac{\nu-2}{\nu}}~ (1-\rho^2)}}$
~~(\ref{kuiigjj})& -   \\ \hline
Gaussian Factor Model & same as (\ref{eq:os_ccc_gv}) & same as
(\ref{eq:os_ccc_gv2}) & same as (\ref{mgjzma}) \\ \hline
Student's Factor Model & $1-{K \over v^2}$~~(\ref{mghhek}) & $1-{K
\over v^2}$~~(\ref{mghhek}) & - \\ \hline
\hline
\end{tabular}
\vspace{5mm}
\caption{\label{table1} Large $v$ and $u$ dependence of the
conditional correlations
$\rho_{v}^+$ (signed condition), $\rho_{v}^s$ (unsigned condition)
and $\rho_{u}$ (on both
variables) for the different models studied in the present paper,
described in the first
column.  The numbers in
parentheses give the
equation numbers from which the formulas are derived. The factor
model is defined by
(\ref{mgmmsaa}), i.e., $X= \alpha Y + \epsilon$. $\rho$ is the
unconditional correlation
coefficient.
}
\end{center}
\end{table}

\begin{table}
\begin{center}
\begin{tabular}{||c|c|c|c|c|c||}
\hline
\hline
& $\rho_{v=\infty}^+$ & $\rho_{v=\infty}^s$ &$\rho_{u=\infty} $&$\lambda$&
\hspace{0.5cm} $\bar \lambda$ \hspace{0.5cm} \\ \hline
Bivariate Gaussian & 0 & 1 & 0 & 0 & $\rho$\\
\hline
Bivariate Student's   & see Table \ref{table1} &  see Table \ref{table1} & -& $
2 \cdot \bar T_{\nu+1} \left(\sqrt{\nu+1} \sqrt{\frac{1-\rho}{1+\rho}} \right)
$& 1 \\ \hline Gaussian Factor Model & 0 & 1 & 0& 0 & $\rho$\\
\hline
Student's Factor Model & 1 & 1 &-& $ \frac{\rho^\nu}{\rho^\nu+ ( 1-\rho^2)
^{\nu/2}}$ & 1 \\
  \hline \hline
\end{tabular}
\vspace{5mm}
\caption{\label{table2} Asymptotic values of
$\rho_{v}^+$, $\rho_{v}^s$ and $\rho_{u}$ for $v \to +\infty$ and $u
\to \infty$ and
comparison with the tail-dependence $\lambda$ and $\bar \lambda$ for the four
models indicated in the
first column. The factor model is defined by
(\ref{mgmmsaa}), i.e., $X= \alpha Y + \epsilon$.
$\rho$ is the unconditional correlation coefficient. For the
Student's factor model, $Y$ and $\epsilon$ have
centered Student's distributions with the same number $\nu$ of
degrees of freedom and
their scale
factors are respectively equal to $1$ and $\sigma$, so that $\rho =
(1+\frac{\sigma^2}{\alpha^2})^{-1/2}$. For the Bivariate Student's 
distribution,
we refer to Table 1 for the constant values of $\rho_{v=\infty}^+$ 
and $\rho_{v=\infty}^s$.}
\end{center}
\end{table}

\newpage

\newpage

\newpage

\newpage

\newpage

\newpage

\newpage

\newpage

\newpage

\newpage

\begin{thebibliography}{123}

\bibitem[Abramovitz and Stegun (1972)]{Abra}  Abramovitz, E. and I.A. Stegun,
1972, Handbook of Mathematical functions (Dover Publications, New York).

\bibitem[Andersen and Sornette (2001)]{Soretal3} Andersen, J.V. and D.
Sornette, 2001,
Have your cake and eat it too: increasing returns while lowering large risks!
Journal of Risk Finance 2, 70-82.

\bibitem[Ang and Bekaert (2000)]{AB00} Ang, A. and G. Bekaert, 2000,
International asset allocation with regime shifts, Working paper.

\bibitem[Ang and Chen (2001)]{AC01} Ang, A. and J. Chen, 2001, Asymmetric
correlations of equity portfolios, Working Paper.

\bibitem[Bhansali and Wise (2001)]{Bhansali} Bhansali, V. and M.B. Wise, 2001,
Forecasting portfolio risk in normal and stressed markets, working paper
(preprint at http://xxx.lanl.gov/abs/nlin.AO/0108022)

\bibitem[Bookstaber (1997)]{Bookstaber} Bookstaber, R., 1997, Global
risk management:
are we missing the point? Journal of Portfolio Management, 23, 102-107.

\bibitem[Boyer et al. (1997)]{Boyer_etal} Boyer, B.H., M.S Gibson and M.
Lauretan, 1997, Pitfalls in tests for changes in correlations, International
Finance Discussion Paper 597, Board of the Governors of the Federal Reserve
System.

\bibitem[Davis et al. (1999)]{Davis_etal} Davis, R.A., T. Mikosch and
B. Basrak, 1999, Sample ACF of multivariate stochastic recurrence
equations with application to GARCH, Working paper.

\bibitem[Cizeau et al. (2001)]{Cizeau_etal} Cizeau, P., M. Potters and J.P.
Bouchaud, 2001, Correlation structure of extreme stock returns, Quantitative
Finance 1, 217-222.

\bibitem[Coles et al. (1999)]{Coles_etal} Coles, S., J. Heffernan and J. Tawn,
1999, Dependence measures for extreme value analyses, Extremes 2,
339-365.

\bibitem[Embrechts et al. (1999)]{EMN99}  Embrechs, P., A.J. McNeil and
D. Straumann, 1999, Correlation~: Pitfalls and Alternatives. Risk,
69-71.

\bibitem[Embrechts et al. (2001)]{EMS01} Embrechts, P., A.J. McNeil and
D. Straumann, 2001, Correlation and Dependency in Risk Management~: Properties
and Pitfalls, in : Dempster, M., ed., Value at Risk and Beyond (Cambridge
University Press).

\bibitem[Forbes and Rigobon (2001)]{FR01} Forbes, K.J. and  R. Rigobon, 2001,
No contagion, only interdependence: measuring stock market co-movements,
forthcoming Journal of Finance.

\bibitem[Frees and Valdez (1998)]{FV98} Frees, E. and E. Valdez, 1998,
Understanding Relationships using copulas, North Americam Actuarial
Journal 2, 1-25.

\bibitem[Hartmann et al.  (2001)]{H_etal} Hartman, P., S. Straetmans
and C.G. de
Vries, 2001, Asset market linkages in crisis periods, European Central Bank,
Working paper n$^o$ 71.

\bibitem[Hauksson et al. (2001)]{Muller}
Hauksson, H.A., M.M. Dacorogna, T. Domenig, U.A. M\"uller  and
G. Samorodnitsky, 2001, Multivariate Extremes, Aggregation and Risk Estimation,
Quantitative Finance 1, 79-95.

\bibitem[Heffernan (2000)]{H00} Herffernan J.E., 2000, A directory of tail
dependence, Extremes 3, 279-290.

\bibitem[Hult and Lindskog (2001)]{HL01} Hult, H. and F. Lindskog, 2001,
Multivariate extremes, aggregation and dependence in elliptical distributions,
Risklab working paper.

\bibitem[Jensen (1995)]{J95} Jensen, J.L., 1995, Saddlepoint Approximations
(Oxford University Press).

\bibitem[Joe (1997)]{J97} Joe, H., 1997, Multivariate models and dependence
concepts (Chapman \& Hall, London)

\bibitem[Johnson and Kotz (1972)]{JK72} Johnson, N.L. and S. Kotz, 1972,
Distributions in statistics: Continuous multivariate distributions (John Willey
and Sons).

\bibitem[King and Wadhwani (1990)] {KW90} King, M. and S. Wadhwani, 1990,
Transmission of volatility between stock markets, The Review of Financial
Studies 3, 5-330.

\bibitem[Kulpa (1999)]{K99} Kulpa, T., 1999, On approximations of copulas,
international Journal of Mathematics and Mathematical sciences 22, 259-269.

\bibitem[Ledford and Tawn (1996)]{LT96} Ledford, A.W. and J.A. Tawn,1996
Statistics for near independence in multivariate extrem values,
Biometrika 83, 169-187.

\bibitem[Ledford and Tawn (1998)]{LT98} Ledford, A.W. and J.A. Tawn,
1998, Concomitant tail behavior for extremes, Adv. Appl. Prob. 30,
197-215.

\bibitem[Li et al. (1998)]{Li_etal} Li, X., P. Mikusincki and M.D. Taylor,
1998, Strong approximation of copulas, Journal of Mathemetical Analisys and
Applications 225, 608-623.

\bibitem[Lindskog (1999)]{L99} Lindskog, F., Modelling Dependence
with Copulas, Risklab working paper.

\bibitem[Longin and Solnik (1995)]{LS95} Longin F. and B. Solnik, 1995, Is the
correlation in international equity returns constant: 1960-1990? Journal of
International Money and Finance 14, 3-26.

\bibitem[Longin and Solnik (2001)]{LS01} Longin F. and B. Solnik, 2001,
Extreme Correlation of International Equity Markets, The Journal of Finance
LVI, 649-676.

\bibitem[Loretan (2000)]{Loretan} Loretan, M., 2000,
Evaluating changes in correlations during periods of high market volatility,
Global Investor 135, 65-68.

\bibitem[Loretan and English (2000)]{Loretanenglish}
Loretan, M. and W.B. English, 2000,
Working paper 000-658, Board of Governors of the Federal Reserve
System

\bibitem[Malevergne and Sornette (2001)]{MS01} Malevergne, Y. and
   D.Sornette, 2001, Testing the Gaussian copula hypothesis for
   financial assets dependence, Working paper.

\bibitem[Malevergne and Sornette (2002)]{MS02} Malevergne, Y. and
   D.Sornette, 2002, Tail dependence for factor models, Working paper.y

\bibitem[Mansilla (2001)]{Mansilla} Mansilla, R., 2001,
Algorithmic complexity of real financial markets, Physica A 301,
483-492.

\bibitem[Meerschaert and Scheffler (2001)]{Meerschaert} Meerschaert, M.M. and
H.P. Scheffler, 2001, Sample cross-correlations for moving averages with
regularly varying tails, Journal of Time Series Analysis 22, 481-492.

\bibitem[Nelsen (1998)]{Nelsen} Nelsen, R.B., 1998, An Introduction to
Copulas. Lectures Notes in statistic 139 (Springer Verlag, New York).

\bibitem[Patton (2001)]{P01} Patton, J.A., 2001, Estimation of copula models
for time series of possibly different lengths, U of California, Econ. Disc.
Paper No. 2001-17.

\bibitem[Quintos (2001)]{Quintos1} Quintos, C.E., 2001, Estimating
tail dependence and
testing for contagion using tail indices, working paper.

\bibitem[Quintos et al. (2001)]{Quintos2} Quintos, C.E., Z.H. Fan and
P.C.B. Phillips, 2001,
Structural change tests in tail behaviour and the Asian crisis,
Review of Economic Studies 68, 633-663.

\bibitem[Ramchand and Susmel (1998)]{RS98} Ramchand, L. and R. Susmel, 1998,
Volatility and cross correlation across major stock markets, Journal of
Empirical Finance 5, 397-416.

\bibitem[Ross (1976)]{R76} Ross, S., 1976, The arbitrage theory of capital
asset pricing, Journal of Economic Theory 17, 254-286.

\bibitem[Scaillet (2000)]{S00} Scaillet, O. 2000, Nonparametric
estimation of copulas for time series, Working paper.

\bibitem[Sharpe (1964)]{S64} Sharpe, W., 1964, Capital assets prices: a theory
of market equilibrium under conditions of risk, Journal of Finance, 19,
425-442.

\bibitem[Silvapulle and Granger (2001)]{Granger} Silvapulle, P. and
C.W.J. Granger, 2001,
Large returns, conditional correlation and portfolio diversification:
a value-at-risk
approach, Quantitative Finance 1, 542-551.

\bibitem[Sornette et al. (2000a)]{Soretal1} Sornette, D. P. Simonetti
and J. V. Andersen, 2000,
$\phi^q$-field theory for Portfolio optimization: ``fat tails'' and
non-linear correlations, Physics Report 335, 19-92.

\bibitem[Sornette et al. (2000b)]{Soretal2} Sornette, D., J.V. Andersen
and P. Simonetti, 2000,
Portfolio Theory for ``Fat Tails'',
International Journal of Theoretical and Applied Finance 3, 523-535.

\bibitem[Starica (1999)]{Star99} Starica, C., 1999, Multivariate extremes for
models with constant conditional correlations, Journal of Empirical Finance 6,
515-553.

\bibitem[Tsui and Yu (1999)]{TY99} Tsui, A.K. and Q. Yu, 1999, Constant
conditional correlation in a bivariate GARCH model: evidence from the stock
markets of China, Mathematics and Computers in Simulation 48, 503-509.

\end{thebibliography}
\end{document}